\documentclass{aa}  
\usepackage{xcolor}
\usepackage{appendix}
\usepackage{graphicx}
\usepackage{hyperref}
\usepackage{longtable}
\usepackage{txfonts}
\begin{document}

   \title{Th/Eu abundance ratio of red giants in the Kepler field}

   \subtitle{}

   \author{Ainun Azhari\inst{1, 2, 3}
          \and
          Tadafumi Matsuno\inst{3, 4}
          \and
          Wako Aoki\inst{1, 2}
          \and
          Miho N. Ishigaki\inst{1, 2, 5}
          \and
          Eline Tolstoy\inst{3}
          }

   \institute{Astronomical Science Program, Graduate Institute for Advanced Studies, SOKENDAI, 2-21-1 Osawa, Mitaka, Tokyo 181-8588, Japan
        \\
        \email{ainun-nahdhia.azhari@grad.nao.ac.jp, aoki.wako@nao.ac.jp, miho.ishigaki@nao.ac.jp}
        \and National Astronomical Observatory of Japan, 2-21-1 Osawa, Mitaka, Tokyo 181-8588, Japan
        \and Kapteyn Astronomical Institute, University of Groningen, PO Box 800, 9700AV Groningen, the Netherlands
        \\
        \email{etolstoy@astro.rug.nl}
        \and Astronomisches Rechen-Institut, Zentrum für Astronomie der Universität Heidelberg, Mönchhofstraße 12-14, 69120 Heidelberg, Germany
        \\
            \email{matsuno@uni-heidelberg.de}
        \and Kavli Institute for the Physics and Mathematics of the Universe (WPI), The University of Tokyo Institutes for Advanced Study, The University of Tokyo, Kashiwa, Chiba 277-8583, Japan\\
        }

   \date{}

 
  \abstract{
   The r-process production in the early universe has been well constrained by the extensive studies of metal-poor stars. 
   However, the r-process enrichment in the metal-rich regime is still not well understood. 
   In this study, we examine the abundance ratios of Th and Eu, which represent the actinides and lanthanides, respectively, for a sample of metal-rich disk stars.
   Our sample covers 89 giant stars in the Kepler field with metallicities $-0.7 \leq \rm{[Fe/H]} \leq 0.4$ and ages from a few hundred Myr to $\sim 14$ Gyr.
   Age information for this sample is available from stellar seismology, which is essential for studying the radioactive element Th.
   We derive Th and Eu abundances through $\chi^2$ fitting of high-resolution archival spectra ($R \sim 80,000$) observed with the High Dispersion Spectrograph (HDS) at the Subaru Telescope. 
   We create synthetic spectra for individual stars using a 1D LTE spectral synthesis code, Turbospectrum, adopting MARCS model atmospheres. 
   Our study establishes the use of a less extensively studied Th II line at 5989 \AA, carefully taking into account the blends of other spectral lines to derive the Th abundance.
   We successfully determine Eu abundance for 89 stars in our sample and Th for 81 stars. 
   For the remaining 8 stars, we estimate the upper limits of Th abundance. 
   After correcting the Th abundance for the decay, we find no correlation between $\rm{[Th/Eu]}$ and $\rm{[Fe/H]}$, which indicates that actinides production with respect to lanthanides does not depend on metallicity. 
   On the other hand, we find a positive correlation of $\rm{[Th/Eu]}$ with age, with a slope of $0.10 \pm 0.04$. 
   This may hint at the possibility that the dominant r-process sources are different between the early and late universe.
   }


   \maketitle
%

\section{Introduction}
Heavy elements ($Z \geq 30$) in the universe are produced by neutron-capture nucleosynthesis, an important part of which is the rapid neutron-capture process, known as the r-process \citep{Burbidge1957, Cameron1957}.
The r-process produces about half of the heavy elements, especially heavy elements with $Z \sim 57 -71$ (the lanthanides), and exclusively elements with $Z \geq 90$ (the actinides) \citep{Seeger1965, Kratz1993, Goriely2001}.
For the r-process to happen, the physical conditions must satisfy sufficiently high neutron density and low electron fraction. 
Astrophysical sites that can host the r-process and reproduce the observed stellar abundance pattern are still under debate \citep{Arnould2007, Kajino2019, Cowan2021}.
Among the proposed events are mergers of binary neutron stars (NSM) \citep{Radice2018, Domoto2021, Naidu2022, Holmbeck2023}, neutron star-black hole (NS-BH) mergers \citep{Surman2008, Wehmeyer2019, Wanajo2024}, and highly energetic cases of supernovae such as magnetorotational supernovae (MRD SNe) \citep{Winteler2012, Cescutti2014, Nishimura2017, Reichert2023}.
To date, NSM is the only event proven to host the r-process, observed from the electromagnetic counterpart of the gravitational wave detection \citep{Watson2019, Siegel2022, Kasliwal2022, Domoto2025}.

The abundances of r-process elements have been extensively studied for metal-poor stars, which would have been polluted by a single r-process event \citep{Truran1981, Sneden1996, Farouqi2022}. 
The abundance pattern of metal-poor stars thus can be used to infer the r-process mechanisms and the astrophysical events and their physical properties. 
Europium, as an element with almost pure r-process contribution \citep{Bisterzo2014}, is usually used as a proxy for r-process study in general. 
On the other hand, Th is used as a proxy for actinides, as it has the isotope with the longest lifetime (14 Gyr).
Another long-living actinide, U, is very difficult to detect as the absorption lines reside in the UV part of the stellar spectra and are very weak, often severely blended \citep{Cayrel2001, Frebel2007, Placco2017, Shah2023}.
The studies of metal-poor stars revealed that, while the elements from the second to the third r-process peak, including the lanthanides follow a universal pattern \citep{Sneden1996, Sneden2000, Westin2000, Honda2006}, the actinides show variations where some stars exhibit overabundance or underabundance relative to the scaled solar pattern, termed actinide-boost and actinide-deficient stars, respectively \citep{Cayrel2001, Hill2002, Honda2004, Placco2023}. 
It is still debated whether this variation is caused by different r-process events or a variation in the same event \citep{Schatz2002, Holmbeck2019, Wanajo2024}. 
This highlights the importance of observing Th and Eu abundances, as they can help understand the origin of the r-process production, especially the difference between the lanthanides and actinides.

The understanding of actinide production is crucial to the application of Th as an age indicator.
The $[\rm{Th/Eu}]$ abundance ratios have been used as a chronometer to determine stellar ages \citep{Sneden1996, Cowan1997, Honda2004, Frebel2007}, as an alternative to isochrone methods, which are dependent on stellar evolution models. 
The caveat of using chronometers is the uncertainty of the theoretical initial production ratio (PR) of the elements involved, which contributes to the derived age uncertainty \citep{Cayrel2001, Frebel2007}.
Once the PR is better constrained, the Th/Eu chronometer can provide more accurate age estimates.

While the studies of metal-poor stars give valuable insights into the r-process production in the early universe, the r-process enrichment in the more metal-rich region is less well understood. 
In contrast with metal-poor stars that are enriched by a single or minimum number of r-process sources, metal-rich stars contain a build-up of materials from several generations of star formation and r-process events \citep{Reichert2021}. 
The r-process event progenitors may evolve with time, resulting in different events dominating the metal-rich and metal-poor regimes. 
Thus studying the r-process abundance in metal-rich stars is equally important to provide a complete picture of the r-process enrichment at a later stage and the evolution of the r-process with time. 

In studying metal-rich stars, it is important to disentangle the r-process source(s) from the sources of other elements such as Fe and the $\alpha$-elements.
Previous studies of r-process abundance in metal-rich stars have found that the r-process and the $\alpha$-elements show a different evolution, inferring different properties of supernovae producing them \citep{Guiglion2018}.
The number of contributing sources is also debated,  in the context of the timescale needed to produce the observed r-process abundance in metal-poor and metal-rich stars. 
Several studies found that two sources may be required to explain the observed abundance, one prompt source (e.g., MRD SNe) and one with longer delay time (e.g., NSM) \citep{Wehmeyer2015, Cote2019, Kobayashi2020, Skuladottir2020, Tsujimoto2021, Molero2023}.

The origins of the r-process in disk stars are usually traced by the abundance of Eu, while the actinides are less widely studied.
Recently, Th enrichment in disk stars has been studied by \citet{Mishenina2022}, which compared the $\rm{[Th/Fe]}$ abundance trend with metallicities to the prediction of galactic chemical evolution.
In studies of solar-type metal-rich stars as \citet{Mishenina2022}, Th abundance is derived from the Th line at 4019 \AA\ \citep{delPeloso2005, Botelho2019}.
This line is widely used to determine the Th abundances due to its detectability in solar-type stars.
However, there are difficulties in deriving Th abundance from this line.
One is that the contamination of other atomic and molecular transitions occupy the region around this line.
Some of them just overlap the Th line, e.g., Fe, Ni, Ce, Nd, Mn, and Co, with some weaker contributions from CH and CN.
The Th abundance derived from this line is very sensitive to the estimate of the line strengths of these elements, especially for major blending components such as Fe, Ni, and Mn \citep{Botelho2019}.
Therefore, abundance analysis derived from a Th line with less severe contamination is desired to derive more accurate Th abundances. 

An alternative approach is to utilize a Th line at 5989 \AA, which is less blended compared to the one at 4019 \AA. 
This line is known to provide a consistent Th abundance with those derived from other Th lines for an r-process-enhanced (RPE) very metal-poor giant star, HD 221170 \citep{Ivans2006}.
While this line is weak in non-RPE stars, it has been detected in the spectrum of Arcturus by \citet{Gopka2007} and further confirmed by \citet{Strassmeier2018}.
In such stars, the reliability of the Th abundance derived from this line has yet to be studied. 

In this study, we investigate Th and Eu abundance evolutions across stellar age and metallicities. 
For this purpose, we analyze a sample of red giant stars in the Kepler field, which has prior information on stellar age from asteroseismology, allowing us to apply abundance correction for the decay of Th and study the relation between its initial abundance with age and metallicities.
We focus on analyzing the Th 5989 \AA, examining the reliability of Th abundance measurement from this line.
Our study provides a large sample of metal-rich stars with reliable Th abundances and age information.  

Our study sheds new light on the evolution of Th and Eu in the metal-rich regime. 
We describe the data and methods in Sections \ref{section:data} and \ref{section:methods}.
We present the results and discussion in Sections \ref{section:results} and \ref{section:discussions}, and our conclusions in Section \ref{section:conclusions}.

\section{Data}\label{section:data}
Our sample consists of 89 giant stars, which is a combined sample from \citet{Takeda2015} and \citet{Takeda+(2016)}. 
The sample has seismic parameters (maximum frequency $\nu_{\text{max}}$ and large separation frequency $\Delta \nu$) from NASA's Kepler mission \citep{Koch2010, Mosser2012}. 
\citet{Takeda2015} determined the radii and masses using asteroseismic scaling relations with the seismic parameters determined by \citet{Mosser2012}.
Subsequently, \citet{Takeda+(2016)} determined the age by assigning a stellar evolution model given the seismic mass and radius for each star.
\citet{Liu2019} analyzed the same sample and determined the individual abundances of Na, Al, Mg, Si, Ca, Ti, V, Cr, Ni, Y, Ba, La, and Ce by comparing the observed equivalent widths with theoretical expectations based on the LTE model atmospheres by \citet{Kurucz1993}.
They also determined the Eu abundance for 81 stars in the sample through the equivalent width (EW) method.

In this study, we derive our abundances using high-resolution spectra obtained by \citet{Takeda2015} and \citet{Takeda+(2016)}.
The data were originally taken on September 9th, 2014, and July 3rd, 2015 using the High Dispersion Spectrograph (HDS, \citet{Noguchi2002}) on the Subaru Telescope.
The wavelength coverage is 5100 - 7800 \AA, with a resolving power of $R \sim 80,000$ and signal-to-noise ratio (S/N) of $\sim 100 - 200$.
We retrieved the reduced spectra from the Japanese Virtual Observatory portal for archive HDS spectra\footnotemark{}.
We adopt the stellar parameters ($T_{\rm{eff}}$, $\log g$, and $[\rm{Fe/H}]$) derived by \citet{Takeda2015} and \citet{Takeda+(2016)}.
They determined the stellar parameters by requiring the Fe abundances from individual Fe I and Fe II lines to show no dependence on excitation potentials and equivalent widths, as described in \citet{Takeda2002} and \citet{Takeda2005}.
We adopt elemental abundances reported by \citet{Liu2019} in our analysis, except for Eu.

\footnotetext{\url{http://jvo.nao.ac.jp/portal/subaru/hds.do}}

\section{Methods}\label{section:methods}
We derive the abundance ratios of Eu and Th through spectral synthesis methods based on model atmospheres. 
We create synthetic spectra with TSFitPy \citep{Jeffrey2023, Storm2023}, the Python wrapper for Turbospectrum spectral synthesis code \citep{Alvarez1998, Plez2012} with model atmospheres from an interpolation of MARCS model atmospheres \citep{Gustafsson2008}, assuming LTE conditions and 1D approximation (spherical symmetries for stars with low surface gravities and plane-parallel for stars with high surface gravities).
We determine the best-fit abundance by $\chi^2$ minimization with the Nelder-Mead method, with the $\chi^2$ calculated as:
\begin{equation}
    \chi^2 = \sum_{i=1}  \left( \frac {O_i - E_i}{1/\text{SNR}} \right)^2
\end{equation}
where $O_i$ is the data point and $E_i$ is the synthetic spectrum flux at the corresponding wavelength.
Signal-to-noise ratio (SNR) is defined from the noise around the continuum level near the analyzed Th and Eu lines. 

We analyze Arcturus as a standard star, adopting stellar parameters $T_{\rm{eff}} = 4286 \pm 30$ K, $[\rm{Fe/H}] = -0.52 \pm 0.04$, $\log{g} = 1.66 \pm 0.05$, and $v_{\rm{mic}} = 1.74$ $\rm{kms^{-1}}$ as determined by \citet{Ramirez2011}.
We determine the Eu abundance using the Eu II line at 6645 \AA, taking into account the hyperfine splitting (HFS) already included in the line list described in Section \ref{subsec:linelists}.
The Th abundance is determined using the Th II line at 5989 \AA.
We adopt abundances of O, Na, Mg, Al, Si, K, Ca, Sc, Ti, V, Cr, Mn, Co, Ni, and Zn from \citet{Ramirez2011}.
Of particular interest are the alpha elements, $[\rm{Si/Fe}] = 0.33$ and $[\rm{Mg/Fe}] = 0.37$, whose enhancement contributes to the continuum absorption.
We also adopt typical Nd abundance for disk stars, $[\rm{Nd/Fe}] = 0.2$ \citep{Mishenina2013, Tautvaisien2021} in the calculation of synthetic spectra for the Th line region. 
Since this Nd line is well separated from the Th line, it does not impact our Th measurement, although this Nd abundance results in a better fit with the observation.
We use this star as a reference to adjust the line list as we describe in Section \ref{subsec:linelists}, and apply the adjustment to analyze our main sample.

For the main sample, we adopt the stellar parameters $T_{\rm{eff}}$, $\log{g}$, and $[\rm{Fe/H}]$, estimated by \citet{Takeda2015} and \citet{Takeda+(2016)}.
We also take into account individual elemental abundances from \citet{Liu2019} as mentioned above, and adopt $[\rm{Nd/Fe}] = 0.2$ for the whole sample.
We adopt the solar abundances by \citet{Asplund2009} to obtain the scaled solar abundances of Eu and Th.

\subsection{Atomic and molecular line lists}\label{subsec:linelists}
We use atomic line lists provided with the Turbospectrum code, based on the Gaia-ESO line list \citep{Ryabchikova2015, Heiter2021} with updated atomic data for C, N, O, Mg, and Si \citep{Magg2022}.
The oscillator strength, $\log gf$ value of the Th line adopted in this study is taken from the measurements of \cite{Nilsson2002}.
They reported a total uncertainty of 10\% of the $gf$-value, which translates it to uncertainties of $\sim 0.043$ dex in $\log gf$, and thus the same amount of uncertainties in $A_{\rm{Th}}$.
For comparison, the uncertainty of the $gf$-value for the Th II line at 4019 \AA\ is 3\% ($\log gf = -0.228 \pm 0.013$ \citep{Nilsson2002}).

\begin{figure}[h!]
    \centering
    \includegraphics[width=0.45\textwidth]{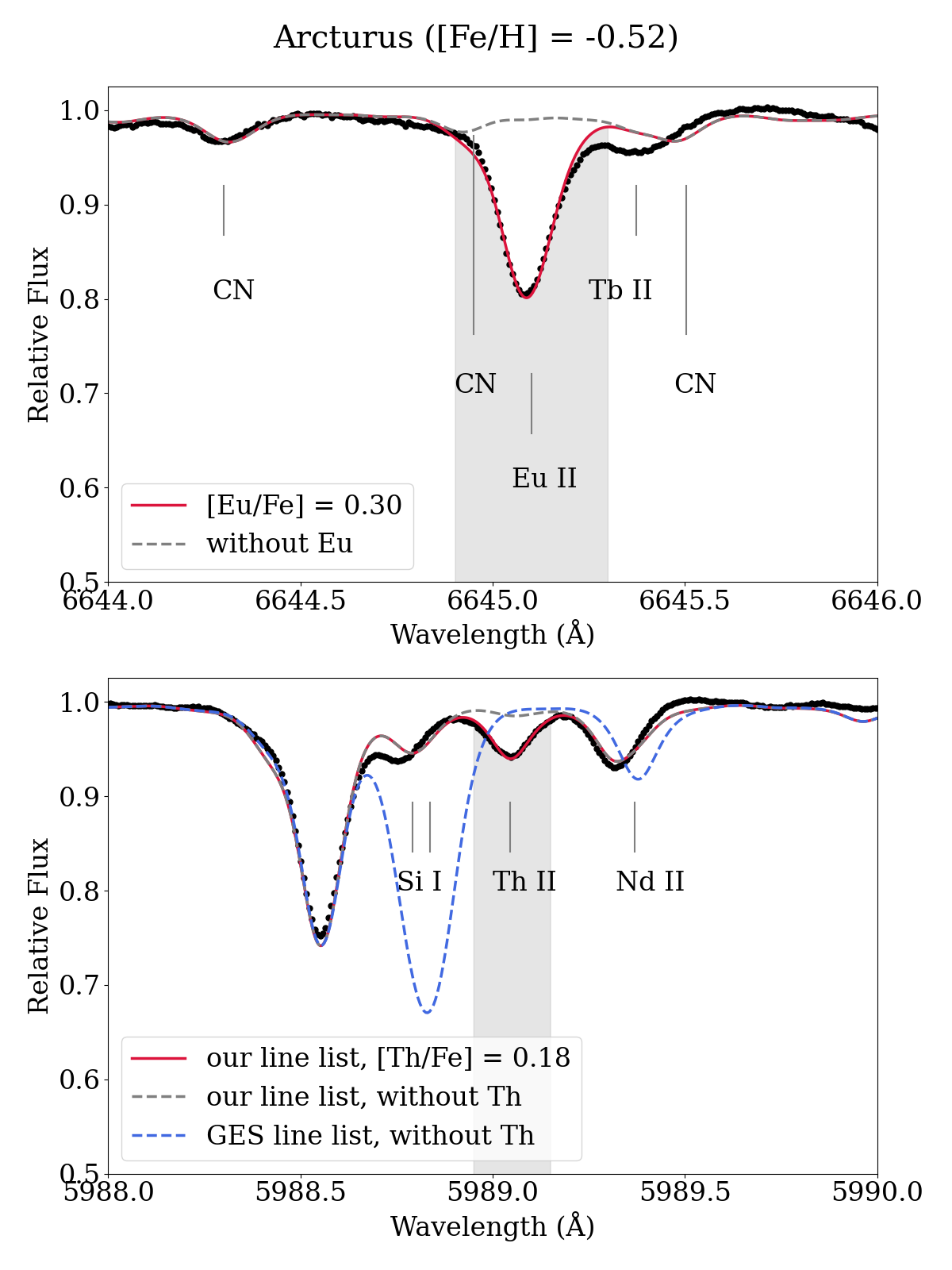}
    
    \caption{\textit{Top panel}: Our fitting result of the Eu line for Arcturus. The black dotted line shows the observed spectrum, the red solid line shows the best-fit spectrum and the gray dashed line shows the spectrum without Eu contribution. 
    \textit{Bottom panel}: Our fitting result of the Th line in Arcturus. The blue dashed lines show the synthetic spectrum before the adjustment of the line list, the red solid line shows the final fitting result, and the gray dashed line shows the synthetic spectrum without Th contribution.}\label{fig:arc_both}
\end{figure}

There are several absorption features around the region of the Th line causing potential blends, i.e., Si I lines toward blue at 5988.791 \AA\ and 5988.838 \AA\ (Figure \ref{fig:arc_both}), an Nd II line toward red at 5989.378 \AA, and a CN line overlapping with the Th line at 5989.054 \AA.
The line list predicts Si and Nd features that do not match the observed feature.
Hence, we make several adjustments to the line list to obtain a better match between the synthetic spectrum and the observation for Arcturus. 

Adopting the Si abundance from \citet{Ramirez2011}, the synthetic spectra of the Si lines are too strong. 
As the Si abundance measurement of \citet{Ramirez2011} is based on the analysis of 5 Si lines, we suspect this discrepancy to be caused by incomplete or incorrect atomic data of the two Si lines.
The oscillator strengths of the two Si lines having the excitation potential of 5.964 eV originate from Robert L. Kurucz's line list. 
As far as we know, there is no information on the reliability of these line data.
As the original line list overestimates the strength of the Si I lines compared to the observed feature, we weaken the lines by gradually changing the oscillator strengths until the synthetic spectrum matches the observation.
We change the oscillator strength from $\log{gf} = -1.453$ to $\log{gf}= -2.053$ for the Si I line at 5988.791 \AA.
The Si I line at 5988.838 \AA\ is neglected.
As for the Nd line, we follow the adjustment made by \citet{Aoki2007}, i.e., the wavelength, oscillation strengths, and excitation potential from 5989.378 \AA\ to 5989.312 \AA, $\log gf = -1.480$ to -2.05, and $\chi$ = 0.745 eV to 0.38 eV.
The final line list we adopt is shown in Table \ref{tab:linelist}.

We also include molecular line lists of 2 isotopes of CN (${}^{12}\rm{C}{}^{14}\rm{N}$ and ${}^{13}\rm{C}{}^{14}\rm{N}$) compiled by \citet{Sneden2014} based on the transition data from \citet{Brooke2014} to fit the Eu line and additional ${}^{12}\rm{C}{}^{12}\rm{C}$ and ${}^{}\rm{TiO}$ molecules from Gaia-ESO lists\footnotemark{} to fit the Th line.
\footnotetext{\url{https://keeper.mpdl.mpg.de/d/6eaecbf95b88448f98a4/?p=\%2Flinelist\%2Fmolecules-420-920nm&mode=list}}
We further inspect the completeness of the CN line lists, especially ${}^{12}\text{C}{}^{14}\text{N}$, which is dominant over the other isotopes. 
In the online line list compiled by R. L. Kurucz\footnotemark{}, two ${}^{12}\text{C}{}^{14}\text{N}$ lines that are absent from the list compiled by \citet{Sneden2014} are found at 5988.438 \AA\ and 5989.054 \AA. 
\footnotetext{R. L. Kurucz online database of molecular line lists, accessible on \url{http://kurucz.harvard.edu/molecules/cn/cnaxbrookek.asc}}
These are the transitions from $J = 51.5$ and $J = 41.5$, respectively, which exceed the range computed by \citet{Brooke2014}.
These lines are also tabulated on the VALD database\footnotemark{}, based on the measurement by Kurucz, from which we confirmed the transition probabilities of the lines. 
\footnotetext{\url{http://vald.astro.uu.se/}}
Of particular interest is the line at 5989.054 \AA\ that overlaps with the Th line we are studying. 
This line has an excitation potential of $\chi = 0.897$ eV and oscillator strength of $\log{gf} = -2.23$, according to Kurucz's online database (see Table \ref{tab:CN_lines}). 

\subsection{CN line strength estimation}\label{subsec:CN_strength_est}
\begin{figure}[h!]
    \centering
    \includegraphics[width=0.45\textwidth]{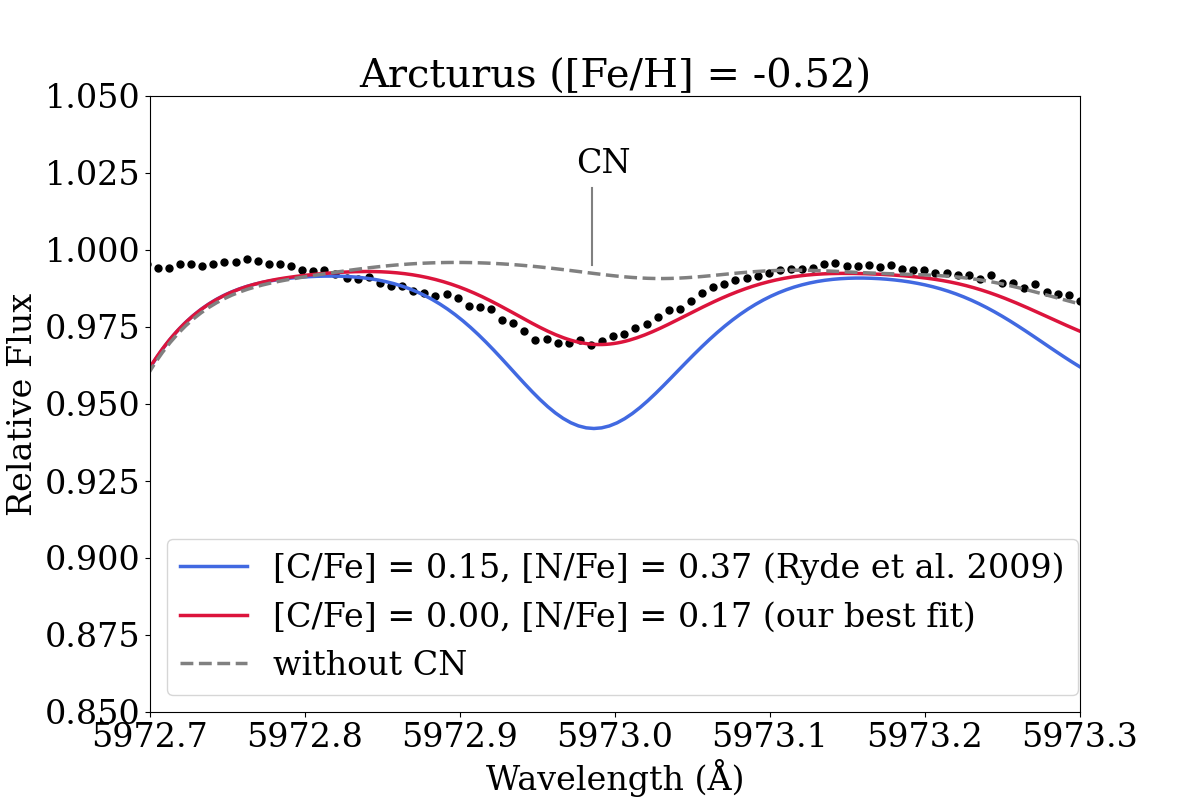}
    
    \caption{CN line we use to constrain the strength of the CN absorption. The blue solid line shows the synthetic spectrum assuming $\rm{[C/Fe]}$ and $\rm{[N/Fe]}$ from \citet{Ryde2009}, while the red solid line shows our best-fit result (see Section \ref{subsec:CN_strength_est}). The gray dashed line shows the synthetic spectrum without CN contribution.}\label{fig:arc_CN}
\end{figure}

We examine the observed features of the aforementioned CN lines in Arcturus, assuming individual abundances of C and N from \citet{Ryde2009}, $[\rm{C/Fe}] = 0.15$ and $[\rm{N/Fe}] = 0.37$.
Our analysis includes other CN lines belonging to the same vibrational transition of the same electronic system in the nearby regions, and the C and N abundances taken from the literature overestimate the strength of those lines. 
Therefore we adjust the C and N abundances using a ${}^{12}\text{C}{}^{14}\text{N}$ line that is not contaminated by atomic absorption.
The CN line at 5972.985 \AA\ is suitable for this purpose (see Table \ref{tab:CN_lines}).
We synthesize spectra assuming several sets of C and N abundances to match the observed feature, as shown in Figure \ref{fig:arc_CN}.
In this case, assuming $[\rm{C/Fe}] = 0.00$ and $[\rm{N/Fe}] = 0.17$ ($[\rm{C+N/Fe}] = 0.04$), which we determine from the method described below, gives a better match to the observed feature.
This is lower than the combined abundance of $\rm{[(C+N)/Fe]} = 0.20$ presented in \citet{Ryde2009}, which was derived from CN lines in the near-infrared region, while the lines we are dealing with lie in the optical region. 

Based on the above discussion, we decide to independently estimate the 5989.054 \AA\ CN line strength from the 5972.985 \AA\ line instead of adopting C and N abundances from the literature. 
We note that our purpose is not to derive the C and N abundance, but only to constrain the strength of the CN line to estimate the contamination on the Th line.
For this purpose, we fit the 5972.985 \AA\ CN line to determine N abundance while fixing the C abundance at $\rm{[C/Fe]} = 0.00$. 
The C and N abundances assumed in the calculations for Arcturus are presented in Figure \ref{fig:arc_CN}.
We then adopt the best-fit $\rm{[N/Fe]}$ for the fixed $\rm{[C/Fe]}$ in the analysis of the Th line. 

The final result of the adjustments of the Th line is shown in the bottom panel of Figure \ref{fig:arc_both}. 
It is clear that the overall agreement between the synthetic and observed spectra is improved by updating the line list.

The Eu line is also minimally blended with a CN line on the blue wing, as shown in the top panel of Figure \ref{fig:arc_both}.
There are multiple CN lines on this region, though the majority of the lines are very weak, the strongest ones being a ${}^{13}\rm{C}{}^{14}\rm{N}$ line at 6644.965 \AA\ and ${}^{12}\rm{C}{}^{14}\rm{N}$ line at 6644.922 \AA\ according to the \citet{Brooke2014} line list. 
However, the contribution from ${}^{13}\rm{C}{}^{14}\rm{N}$ line is not identified in the current analysis.

\begin{figure}
    \centering
    \includegraphics[width=0.45\textwidth]{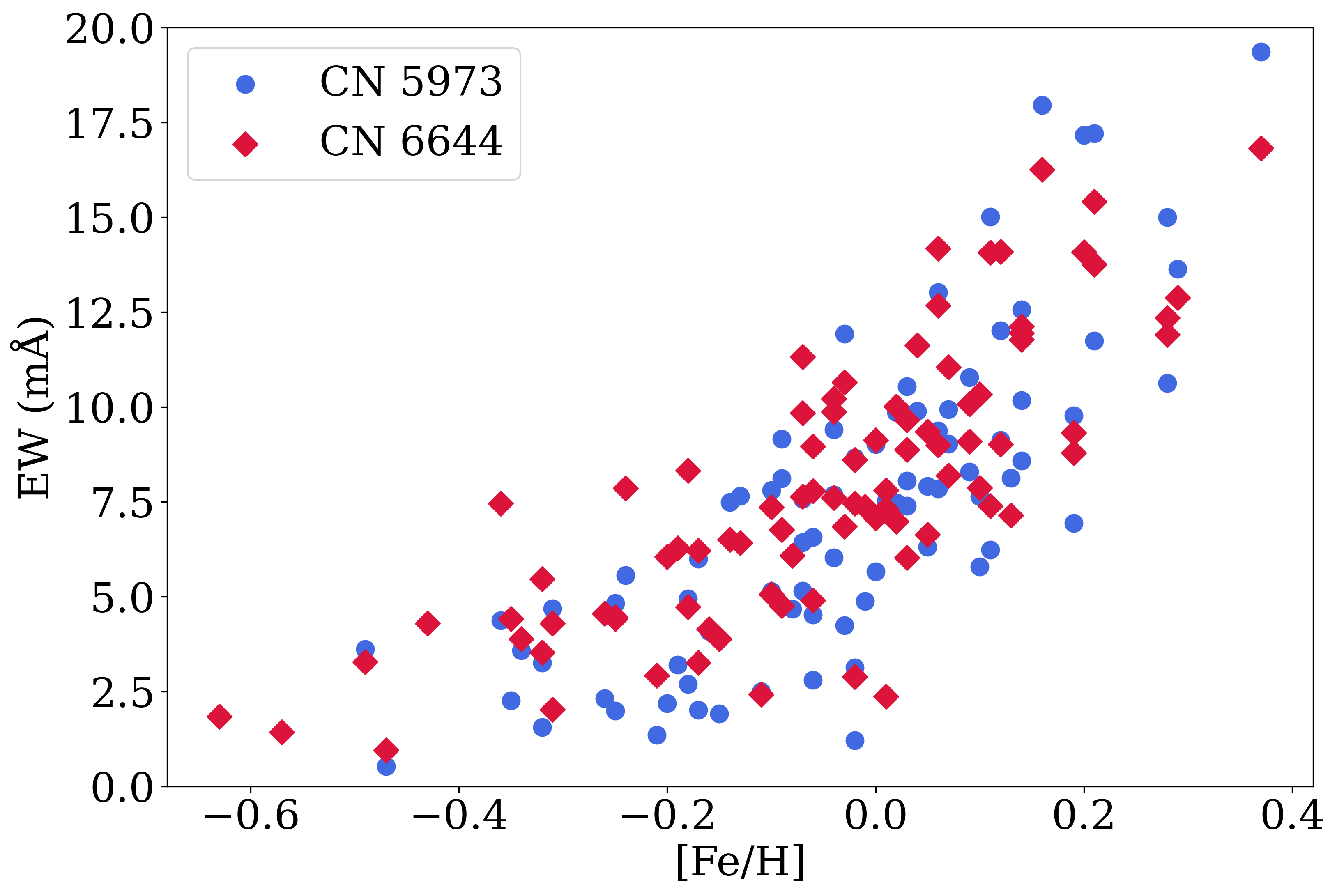}
    
    \caption{Our estimate of the CN line strength (equivalent widths) by fixing the C abundance and varying N abundance. The strength of the CN line varies in the region of the Th and Eu lines, as indicated by the equivalent width of the CN lines at 5982.985 \AA\ and 6644.3 \AA. }\label{fig:N_abund}
\end{figure}

Even though the blend is very weak in the case of Arcturus, the CN line is stronger in the case of more metal-rich stars (see Figure \ref{fig:pick_4_stars_eu}).
We estimate the strength of the CN line at 6644.922 \AA\ using a CN line at 6644.304 \AA\ (overlapping with another line at 6644.352 \AA), which is directly adjacent to the Eu line (see Figure \ref{fig:arc_both}).
The blend on the Eu line slightly affects the Eu abundance determination in cool metal-rich stars, as we describe in Section \ref{section:results}, where we find some discrepancy between our results with that of \citet{Liu2019}.
However, since the blend is located only on the wing of the Eu line, the Eu abundance is not very sensitive to the choice of the CN abundance, as long as the blend itself is included. 
The information on the CN lines used to constrain the strength and the CN lines contaminating the Th and Eu lines is given in Table \ref{tab:CN_lines}.
We apply this method to our main sample.
Figure \ref{fig:N_abund} shows the strength of the CN lines represented by their equivalent widths.
It is evident that the strength of the CN lines increases with metallicities, indicating more prominent blends in more metal-rich stars.

\subsection{Error estimation}
The total error in the Th and Eu abundances stems from the stellar parameters, fitting, continuum placement, and the CN line contribution. 
To estimate the errors due to individual sources, we estimate the sensitivity of the Th and Eu abundances to changes of each parameter.
As the Th line is blended with CN, we need to separate the CN and Th contributions in estimating the errors from stellar parameters, as the change in stellar parameters can also change the CN line contribution.
We also estimate the sensitivity of Th abundances to the change of CN line strengths for a fixed set of stellar parameters.
We apply the same treatment as Th for the Eu line. 

For our main sample, the intrinsic errors of the stellar parameters have been estimated by \citet{Takeda2015} and \citet{Takeda+(2016)} to be $\sim 10-30$ K for $T_\mathrm{eff}$, $\sim 0.05-0.1$ dex for $\log{g}$, $\sim 0.05-0.1$ km $\rm{s^{-1}}$ for microturbulent velocity and 0.02–0.04 dex for $\rm{[Fe/H]}$ \citep{Takeda2008}.
Including the systematic difference from the stellar parameters in the \textit{Kepler} Input Catalogue, we estimated the errors of $T_{\mathrm{eff}}$, $\log{g}$, $v_{\rm{mic}}$ and $\rm{[Fe/H]}$ to be 100 K, 0.1 dex, 0.2 km $\rm{s^{-1}}$ and 0.1 dex, respectively, following \citet{Liu2019}.
For Arcturus, we consider the uncertainties given by \citet{Ramirez2011}, 30 K for $T_{\rm{eff}}$, 0.05 dex for $\log{g}$, and 0.04 dex for $\rm{[Fe/H]}$.
We do not consider the uncertainty from microturbulent velocity in Arcturus as it is not given by \citet{Ramirez2011}. 
The effect of microturbulent velocity is expected to be negligible as we demonstrated for our main sample. 
The typical errors from stellar parameters for Th and Eu are given in Table \ref{tab:err_est}.

Since we determine the abundance by $\chi^2$ fitting, we adopt the fitting error from the $\chi^2$ calculation. 
The abundance offset at 1$\sigma$ is taken from the abundance for which the $\chi^2$ value is different by 1 from the best-fit value.
We estimate continuum placement error by shifting the continuum level by 0.5\% uniformly across the fitting window. 
We find that continuum error is one of the dominant contributors to the total error for Th, along with the fitting error.

We estimate the uncertainty of the strength of the 5989.054 \AA\ CN line from the fitting and the continuum placement error of the 5972.985 \AA\ line.
The change in $\rm{[Th/H]}$ due to the error of the CN line strength varies from 0.0 to 0.1 dex in most cases, and up to $\sim 0.3$ in the most severe cases.
On the other hand, even though the Eu line is blended with a CN line on the blue wing, the effect of the uncertainty of the CN line strength on the Eu abundance is negligible (see Table \ref{tab:err_est}). 

The total error is the sum of the individual sources of errors in quadrature.
The errors of the Eu abundance are smaller than 0.1 dex in most cases, while the errors of Th vary significantly with the strength of the Th line and the SNR.
There is no significant trend between stars with different ranges of stellar parameters for both Eu and Th, as the CN contributions are separated in estimating the errors from stellar parameters. 
We show three example stars with varying Th line strength and SNR to illustrate their impact on the total uncertainty in the Th abundance.
These three stars represent the best, typical, and worst cases regarding the Th line fitting quality.
The total error is dominated by the fitting error, which is dependent on the SNR, and the continuum placement error, as shown in Table \ref{tab:err_est}.

\begin{table*}[htbp]
\centering
\begin{tabular}{llllllllll}
\hline
 (1) & (2) & (3) & (4) & (5) & (6) & (7) & (8) & (9)  & (10) \\
  & $\rm{[X/H]}$ & $\sigma_{\rm{fit}}$ &  $\sigma_{\rm{CN}}$  &  $\sigma_{\rm{cont}}$  & $\sigma_{T_{\rm{eff}}}$  & $\sigma_{\rm{[Fe/H]}}$  & $\sigma_{\rm{\log g}}$ & $\sigma_{\rm{v_t}}$  & $\sigma_{\rm{total}}$ \\
\hline
\hline
main    & Th      & 0.08      & 0.07    & 0.13       & 0.03      & 0.04     &  0.04       & 0.00      & 0.12              \\
sample    & Th      & 0.18      & 0.11    & 0.13       & 0.03       & 0.04     &  0.04       & 0.00      & 0.22              \\
    & Th      & 0.41      & 0.27    & 0.13       & 0.03       & 0.04    &  0.04       & 0.00      & 0.49              \\
  & Eu      & 0.02      & 0.01    & 0.03       & 0.01      & 0.03     & 0.04       & 0.00      & 0.07            \\
\hline
Arcturus & Th     & 0.02      & 0.02    & 0.03   & 0.01       & 0.01      & 0.02   &    & 0.05        \\
   & Eu      & 0.01     & 0.0     & 0.01    & 0.03       & 0.01      & 0.02    &   & 0.04     \\
\hline
\end{tabular}
\vspace{5pt}
\caption{Typical errors for Arcturus and our main sample, each column denotes each uncertainty source: (3) $\chi^2$ fitting, (4) CN line strength, (5) continuum placement, (6) $T_{\rm{eff}}$, (7) $\rm{[Fe/H]}$, (8) $\log g$, (9) microturbulence velocity, and (10) the total error in quadrature.}
\label{tab:err_est}
\end{table*}

\section{Results}\label{section:results}
\subsection{Arcturus}\label{subsec:arcturus}
\begin{figure*}[h!]
    \centering
    \includegraphics[width=1.0\textwidth]{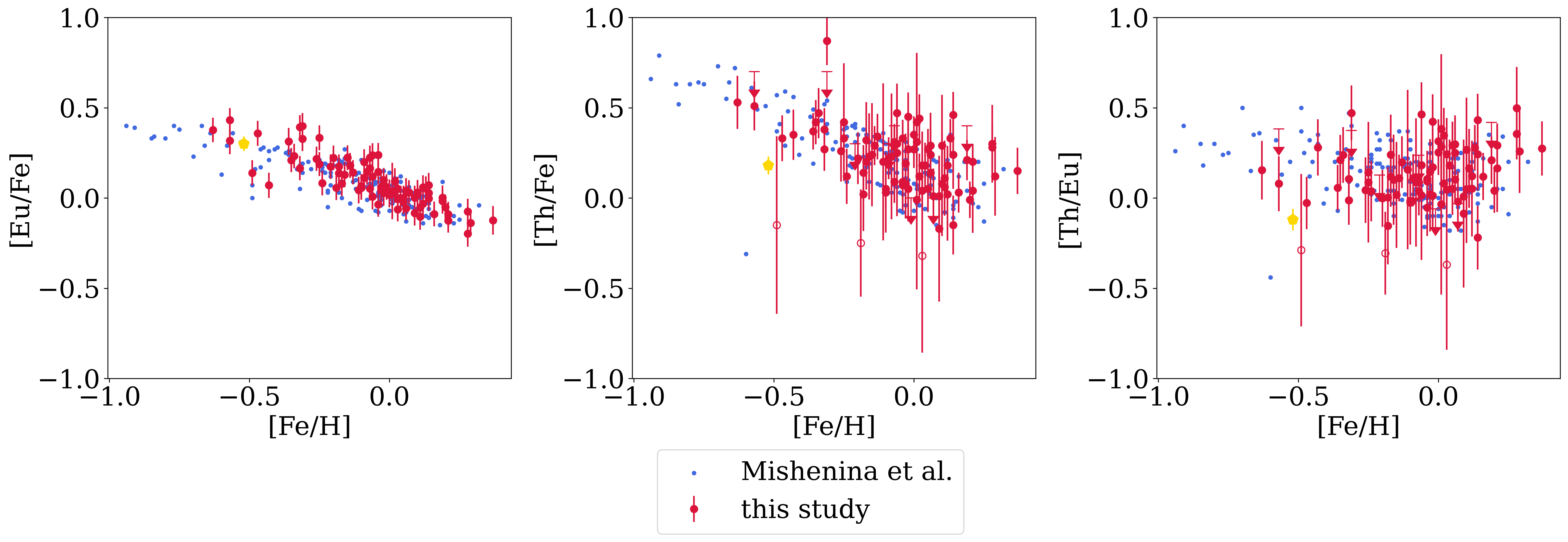}
    
    \caption{From left to right panel: the trend of $[\rm{Eu/Fe}]$, $[\rm{Th/Fe}]$, and $[\rm{Th/Eu}]$ against metallicities, respectively. The blue filled circles show the results of \citet{Mishenina2022}, and the red filled circles show our results. The red open circles show the outliers discussed in the text, and the downward arrows denote upper limits. The yellow pentagon denotes Arcturus.}\label{fig:all_res_feh}
\end{figure*}

The results of Eu and Th abundances for Arcturus are given in Table \ref{tab:redgiants} and shown in Figure \ref{fig:all_res_feh} with those for the main sample.

We compare our results with those of previous studies that use Arcturus as a standard star in their analysis.
\citet{Worley2009} obtained $[\rm{Eu/Fe}] = 0.36 \pm 0.04$, which corrresponds to $\rm{[Eu/H]} = -0.23$, assuming their metallicity of $\rm{[Fe/H]} = -0.59 \pm 0.12$.
\citet{vdSwaelmen2013} found a value of $[\mathrm{Eu/Fe}] = 0.40 \pm 0.05$, corresponding to $\rm{[Eu/H]} = -0.25$ assuming their metallicity of $\rm{[Fe/H]} = -0.65$.
The two literature results are in agreement with our result of $\rm{[Eu/H]} = -0.22 \pm 0.04$ within the uncertainty.

With the updated line list, we obtain $[\mathrm{Th/H}] = -0.34$ or $[\mathrm{Th/Fe}] = 0.18$.
As far as we are concerned, the latest measurement of Th in Arcturus was done by \citet{Gopka1999}, who reported $\log{\epsilon\rm{(Th)}} = -0.31$ from two lines at 4510.526 \AA\ and 4631.762 \AA. 
Adopting the solar Th abundance from \citet{Asplund2009}, this corresponds to $\rm{[Th/H]} = -0.33$, in good agreement with our result.
An earlier study by \citet{Holweger1980} reported a value of $[\mathrm{Th/H}] = -0.90$ obtained using the 4019 \AA\ line.
This result was derived by employing the oscillator strength from \citet{Corliss1979}, $\log{gf} = -0.19$.
A newer measurement by \citet{Nilsson2002} gives a value of $\log{gf} = -0.228$.
With the updated $\log{gf}$, their abundance would correspond to $\rm{[Th/H]} = -0.862$. 
This value is significantly lower than our result.
As the 4019 \AA\ line is more severely affected by blends, the Th abundance derived from this line is very sensitive to the contribution of the blending components.
For example, \citet{Holweger1980} assumed $\log{gf} = -2.60$ for a blend by Co I line at 4019.126 \AA, while a more recent value adopted by \citet{Mishenina2022} is $\log{gf} = -3.298$.
We suspect that the blending of the Co I line is overestimated in their analysis, which leads to an underestimation of the Th abundance.

\subsection{Main sample}\label{subsec:main_sample}
\begin{figure*}
    \centering
    \includegraphics[width=1.0\textwidth]{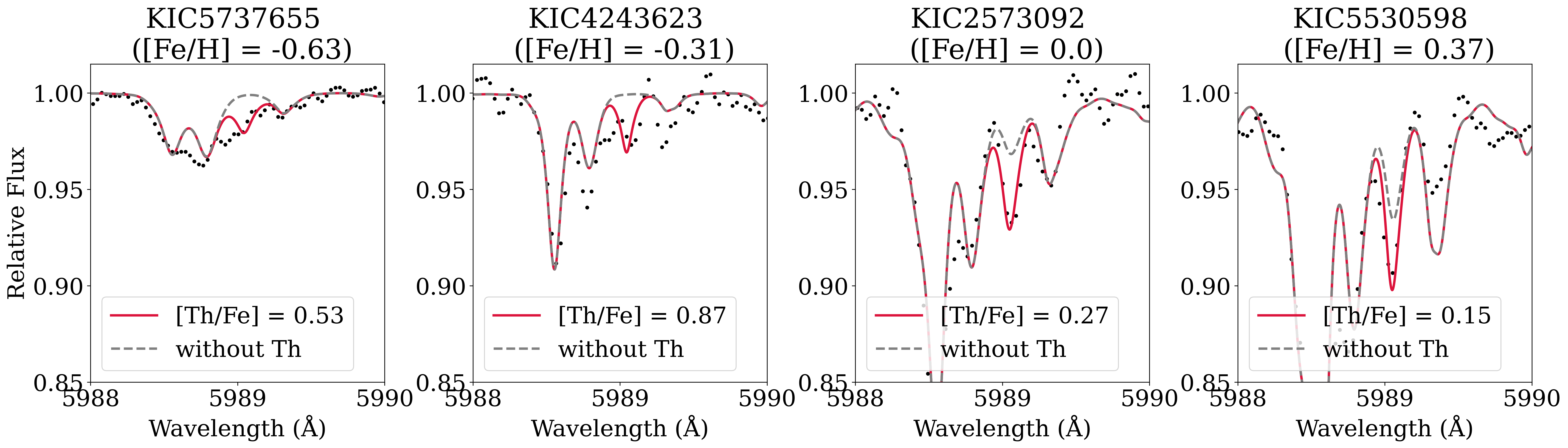}
    
    \caption{Observed spectra of the Th line region (dots) compared with synthetic ones (lines). The fitting results of 4 stars in our sample with different metallicities show varying strength of the CN line contribution. The red solid line shows the best-fit spectrum and the gray dashed line shows the synthetic spectrum without Th contribution (i.e., the contribution from the CN line). The second left panel shows the case of the star with high Th abundance.}\label{fig:pick_4_stars}
\end{figure*}

\begin{figure*}
    \centering
    \includegraphics[width=1.0\textwidth]{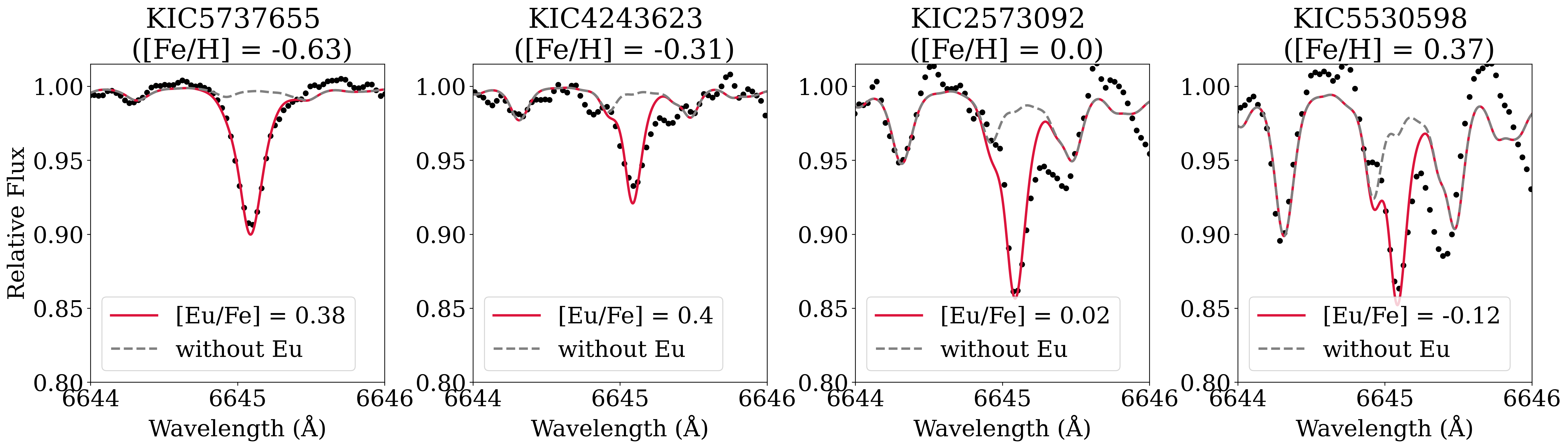}
    
    \caption{Same as Figure \ref{fig:pick_4_stars}, but for the Eu line region.}\label{fig:pick_4_stars_eu}
\end{figure*}

We successfully derive Eu for all the stars in our sample. 
We derive Th abundance for 81 stars in our sample, for which $\chi^2$ fitting reaches convergence. 
For the remaining 8 stars, in which the Th line is very weak or the feature is unclear, we estimate upper limits by searching for Th abundance that gives line depth larger than the noise level. 
We show the best-fit spectra for 4 example stars with different metallicities in Figure \ref{fig:pick_4_stars} and \ref{fig:pick_4_stars_eu}.
The figures show that the contribution of the CN line varies with metallicities, both in the cases of overlap with the Th line and the blue wing of the Eu line.

The Th and Eu abundances for our sample are given in Table \ref{tab:redgiants}.
We show $[\rm{Eu/Fe}]$, $[\rm{Th/Fe}]$, and $[\rm{Th/Eu}]$ against metallicity in Figure \ref{fig:all_res_feh}, along with the results from \citet{Mishenina2022} for comparison.
We note that there are some outliers in our measurements that show low $\rm{[Th/Fe]}$ for their metallicities, KIC10382615 ($\rm{[Th/Fe]} = -0.15$ at $\rm{[Fe/H]} = -0.49$), KIC5858034 ($\rm{[Th/Fe]} = -0.25$ at $\rm{[Fe/H]} = -0.19$), and KIC4902641 ($\rm{[Th/Fe]} = -0.32$ at $\rm{[Fe/H]} = 0.03$).
For KIC10382615, the Th feature is disturbed by noise, and for KIC5858034 and KIC4902641, we noticed a weak absorption feature, with line depth less than 5\%.
This results in large fitting uncertainties for these stars, $\sim 0.4$ dex for KIC10382615 and KIC4902641; and $\sim 0.2$ dex for KIC5858034.
Taking into account the uncertainties, the three stars are not distinguished from the main trend of $[\rm{Th/Fe}]$.

We find one star with a high $[\rm{Th/Fe}]$, KIC4243623 ($[\rm{Th/Fe}] = 0.87$ at  $[\rm{Fe/H}] = -0.31$). 
Since this star shows no discernible CN feature at 5972.985 \AA, we assume no contribution of CN.
Even if we assume an upper limit for the CN contribution, i.e., $\rm{[C/Fe]} = 0.0$ and $\rm{[N/Fe]} = 0.0$, we still find a high value of $[\rm{Th/Fe}] \sim 0.80$ for this star.
We also find a relatively high Eu abundance of $[\rm{Eu/Fe}] = 0.40$.
This star is regarded as an object with a slight enhancement in r-process elements.
The observed and best-fit spectra for this star are shown in the second left panel of Figures \ref{fig:pick_4_stars} and \ref{fig:pick_4_stars_eu}.
\citet{Liu2019} classified this star to be a thin disk star, and it does not show any enhancements in other elements, for example, $\rm{[Mg/Fe]} = 0.18$, $\rm{[Ba/Fe]} = -0.1$, and $\rm{[La/Fe]} = -0.03$. 
This finding is an interesting case where a metal-rich star exhibits such a high enhancement in r-process elements, especially Th (see \citet{Xie2024}).

\subsection{Comparison with previous studies}
\begin{figure}[h!]
    \centering
    \includegraphics[width=0.45\textwidth]{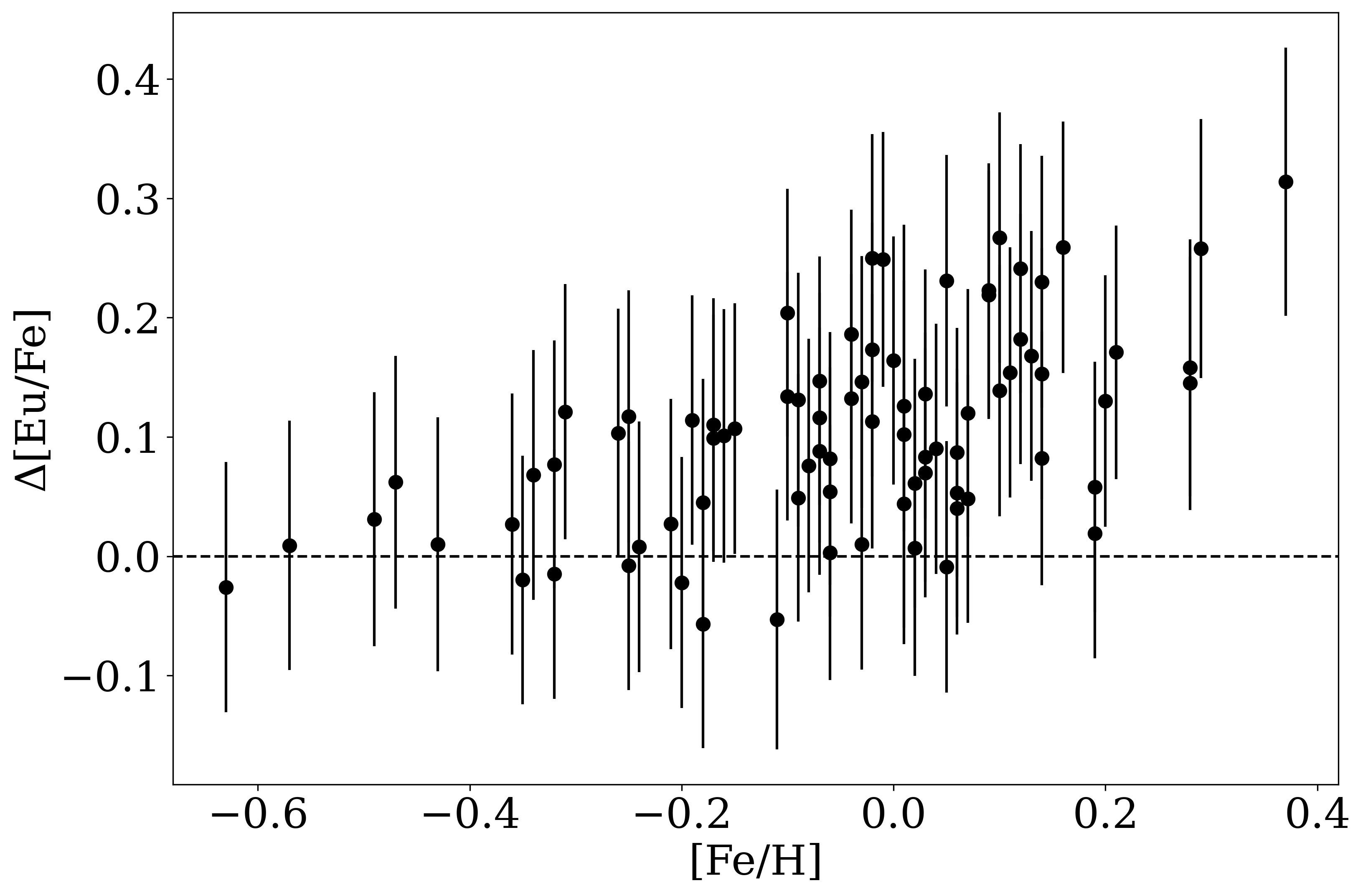}
    
    \caption{Difference between $\rm{[Eu/Fe]}$ in \citet{Liu2019} and this study ($\Delta\rm{[Eu/Fe]} = \rm{[Eu/Fe]}_{\text{literature}} - \rm{[Eu/Fe]}_{\text{this work}}$).}\label{fig:comp_diff}
\end{figure}

We analyze the same sample as \citet{Liu2019}, who also report on $[\rm{Eu/Fe}]$ (see Figure \ref{fig:comp_diff}).
The $\rm{[Eu/Fe]}$ values obtained by our analysis are lower than those from \citet{Liu2019} for the majority of stars in the sample. 
While we derive our abundance based on the spectral synthesis, \citet{Liu2019} relied on the measurement of the equivalent width of the line, in which the blend with CN lines would not be taken into account.
The discrepancy is larger for stars at higher metallicities, suggesting a more significant blend of the CN line.
Although \citet{Liu2019} does not provide any information on the Eu lines they used to derive abundance, we suspect that the treatment of the blend of the CN line is the reason for this offset.
Our findings for $[\rm{Eu/Fe}]$ align with the trend observed by \citet{Mishenina2022}, as shown in the left panel of Figure \ref{fig:all_res_feh}. 

Our $[\rm{Th/Fe}]$ abundance trend is shown in the middle panel of Figure \ref{fig:all_res_feh} along with the results of \citet{Mishenina2022}. 
The overall trend of our $[\rm{Th/Fe}]$ agrees with theirs, albeit with a slightly larger scatter around $[\rm{Fe/H}] \sim 0.0$.
This scatter is also found in the $[\rm{Th/Eu}]$ distribution in the right panel of Figure \ref{fig:all_res_feh}.
We discuss this scatter in more details in Section \ref{section:discussions}.

\begin{figure}[h!]
    \centering
    \includegraphics[width=0.45\textwidth]{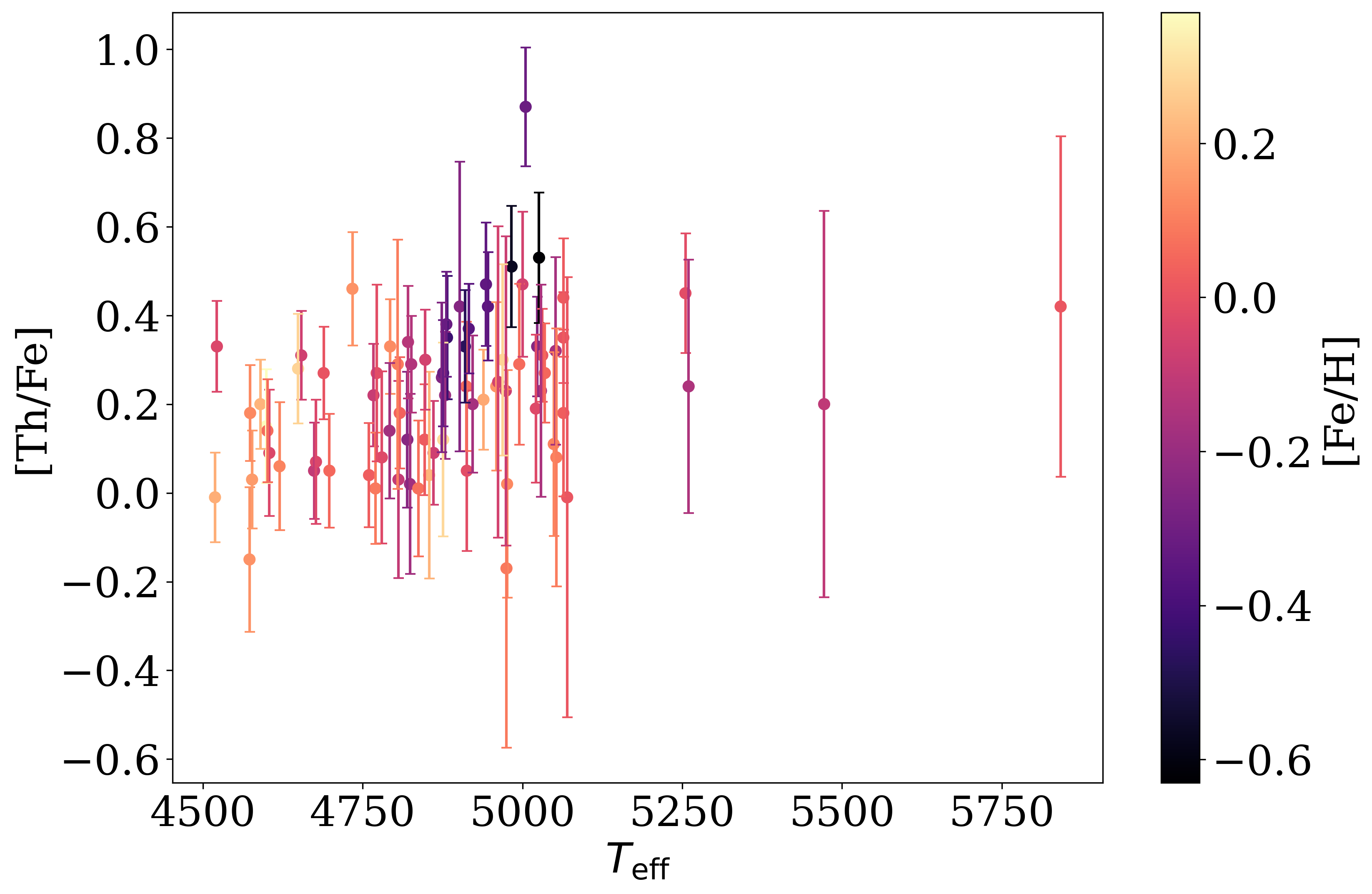}
    
    \caption{$\rm{[Th/Fe]}$ distribution of the stars in our sample, with color codes according to the metallicities. There is no correlation between Th abundance and $T_{\rm{eff}}$ or $\rm{[Fe/H]}$, supporting that our measurement is not affected by systematics in stellar parameters.}\label{fig:teff_hue}
\end{figure}

We note that the \citet{Mishenina2022} determined Th abundances for a sample of main-sequence stars from the 4019 \AA\ Th II line that is significantly affected by the blending of multiple atomic and molecular features, the strongest one being Fe. 
Therefore, some offset could arise between the Th abundances derived from the two lines if the treatment of blending is insufficient. 
Figure \ref{fig:teff_hue} confirms that our $\rm{[Th/Fe]}$ shows no clear correlation with $T_{\rm{eff}}$, supporting that we were able to model the line and blending correctly. 

To further test the consistency of Th abundances derived from both Th lines, we analyze a metal-poor giant star for which both lines are measurable, HD 221170.
We do not find any significant offset between the results from the two lines in this star, with $\rm{[Th/Fe]} = 0.89$ from both lines.
This is in line with the study by \citet{Ivans2006}, which showed an agreement of Th abundances derived from different lines.
They found $A_{\rm{Th}} = -1.49$ from the Th II line at 5989 \AA\ and and $A_{\rm{Th}} = -1.46$ from the line at 4019 \AA\ \citep{Ivans2006}. 
Therefore we conclude that both Th lines give consistent abundances and the results are robust.

It should be noted, however, that comparing Th measurements from both lines as in the study of \citet{Ivans2006} is not possible for metal-rich stars. 
In main-sequence stars, the 5989 \AA\ line is too weak even at solar metallicity, while in metal-rich giants, the 4019 \AA\ line is not measurable because of severe contaminations by other elements.
Our sample and that of \citet{Mishenina2022} consist of stars at higher metallicities, where the effects of molecules can be important, rendering a direct comparison of Th abundance from both lines as in the case of HD 221170.

\subsection{NLTE effects}
Our analysis adopts model atmospheres from MARCS models and the Turbospecrum spectral synthesis code, which assume 1D LTE conditions.
The NLTE effects on Eu abundance have been studied for the 6645 \AA\ line with 1D model atmospheres \citep{Guo2025}.
The correction is expected to be around 0-0.05 dex for red giants in our metallicity range, which is within our Eu abundance uncertainties.  
In the case of Th, \citet{Mashonkina2012} showed that there is a positive correction around 0-0.2 dex in cool stars for the 4019 \AA\ line, though there have not been any NLTE calculations to date for the 5989 \AA\ line.
Although NLTE effects are not expected to be large for ionized lines such as the Th II line we use, future detailed NLTE calculations for this line would be of great interest to investigate the possible effects.

\section{Discussions}\label{section:discussions}
This study presents the first determination of Th abundance from the 5989 \AA\ line, which has not been widely used previously, for a large sample of red giant stars. 
Our study demonstrates that the use of this line offers a new window to derive Th abundance in a larger number of stars.

The advantage of our sample is that ages are estimated by assigning stellar evolution models based on masses and radii derived from asteroseismic scaling relations. 
This enables us to correct the effect of decay of the actinide Th to obtain the initial abundance ratios.
Due to radioactive decay, the currently observed Th abundance does not reflect the initial production of Th.
Thus examining the trend of initial Th abundance is more suitable to accurately infer the correlation between Th production with metallicities and ages.

We correct the Th abundance for the decay by adopting the stellar age provided by \citet{Takeda+(2016)}.
The abundance of Th after some time $t$ will decrease according to the following equation:
\begin{equation}
    N(t) = N_0 \exp{(-\lambda t)}
\end{equation}
In the logarithmic scale of $\rm[{Th/H]}$, the difference between the initial and the current Th abundance is:
\begin{equation}\label{eq:decay_in_log}
    \rm{[Th/H]}_{\rm{initial}} - \rm{[Th/H]}_{\rm{now}} = \frac{\lambda t}{\ln(10)} = 2.15 \times 10^{-2} t/Gyr
\end{equation}
with $t$ the stellar age in Gyr. 
Whereas the age uncertainty is not provided by \citet{Takeda+(2016)}, typical uncertainties from similar methods are 10-20\% for red giants \citep{Yu2018, Warfield2024, Pinsonneault2025}.
This would result in $\sim0.07$ dex change of age in the logarithmic scale, and a 0.02 dex change in the Th abundance correction, which is smaller than our measurement uncertainty. 
Therefore, the impact of age uncertainty on the abundance trend is negligible compared to the uncertainty of abundance measurement.

In the discussion on the Th abundances corrected for the decay ($\rm{[Th/H]}_{\rm{initial}}$ from here onwards), we assume that the r-process material was produced shortly before star formation. 
In reality, there might be some delay between the r-process production and star formation, and Th might have decayed during that period.
Furthermore, accumulated Th from several r-process events might have decayed during the time between subsequent events.

\subsection{$[\mathrm{Eu/H}]$ and $[\rm{Th/H}]$ trends with metallicities and ages}
We first examine the possibility of different enrichment histories between stellar populations in our sample. 
Classifications of thin/thick disk, halo, and transition population have been determined by \citet{Liu2019}.
Thin disk stars dominate our sample, with 5 thick disk, 2 halo, and 2 transition stars in the sample (see Table 4 of \citet{Liu2019}).
The only distinct difference is that the halo stars are of low metallicity, while the thick disk and transition stars overlap with thin disk stars in metallicities. 
Due to this distribution, we find no distinguishable difference in the trends of the abundance ratios between each population with our current sample.

\begin{figure}
    \centering
    \includegraphics[width=0.45\textwidth]{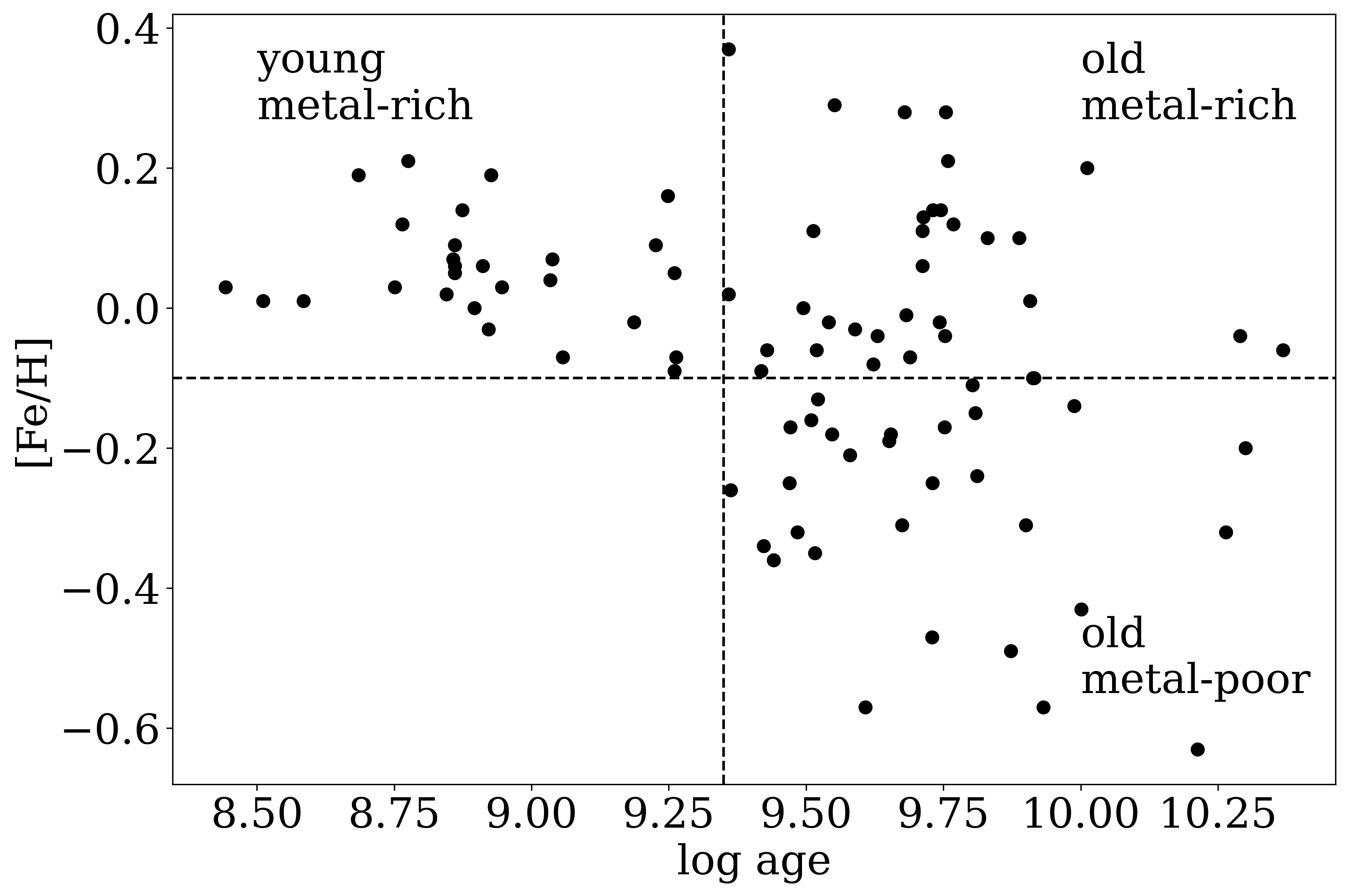}
    
    \caption{Age-metallicity relation for our sample.}\label{fig:age_metal}
\end{figure}

Figure \ref{fig:age_metal} shows that age and metallicity do not exhibit a one-to-one relationship.
Our sample can be divided into three groups, young metal-rich, old metal-rich, and old metal-poor stars, as indicated in the figure.
As our sample covers stars with $\rm{[Fe/H]} \geq -0.7$, there are no strictly "metal-poor" stars, the term is merely used to refer to stars with lower metallicity than the majority of the sample. 
Here we define metal-rich stars as those with $\rm{[Fe/H]} \geq -0.1$ and young stars as those less than $\sim2$ Gyrs of age. 
This is based on the borders of the region with no stars in the sample, located in the bottom left of Figure \ref{fig:age_metal}.

\begin{figure*}[h!]
    \centering
    \includegraphics[width=0.95\textwidth]{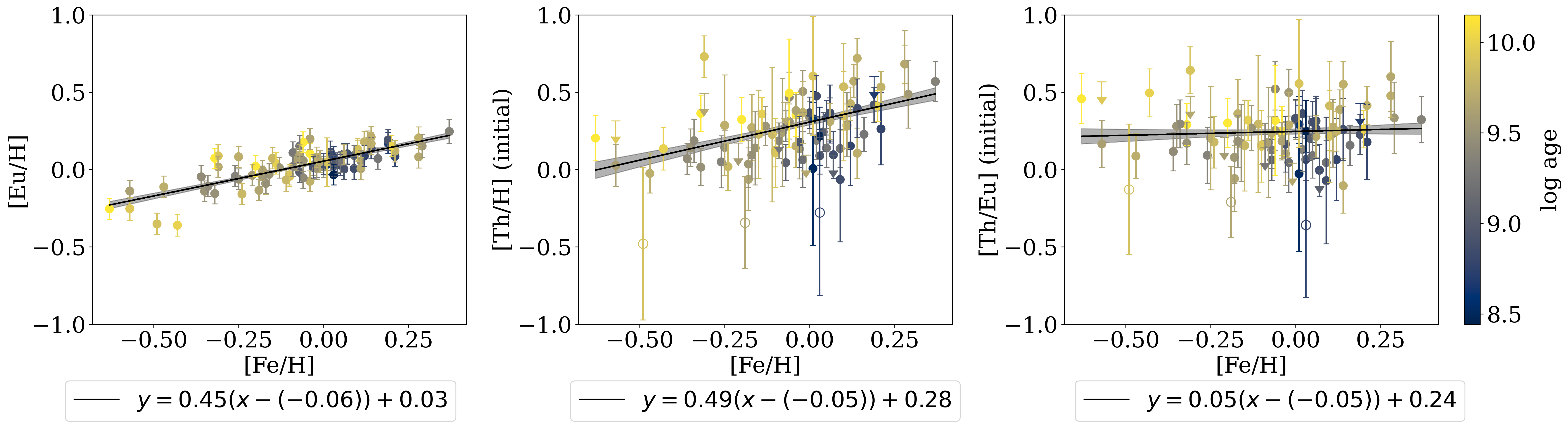}
    
    \caption{Trends of $\rm{[Eu/H]}$, $\rm{[Th/H]}$, and $\rm{[Th/Eu]}$ against metallicity. The symbols are the same as in Figure \ref{fig:all_res_feh}, with the color codes indicating ages. The best-fit linear regression is given below each panel, and the shaded region shows the confidence interval of the fit. We find significant trends for $\rm{[Eu/H]}$ and $\rm{[Th/H]}$, but not for $\rm{[Th/Eu]}$ (see text).}\label{fig:all_res_feh_initial}
\end{figure*}

\begin{figure*}[h!]
    \centering
    \includegraphics[width=0.95\textwidth]{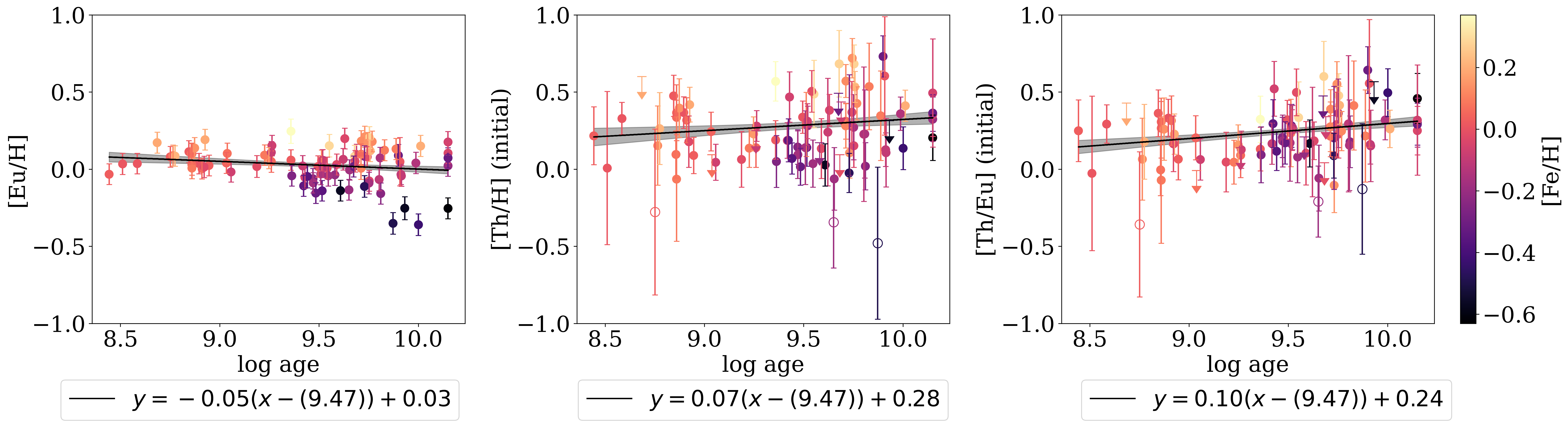}
    
    \caption{Trends of $\rm{[Eu/H]}$, $\rm{[Th/H]}$, and $\rm{[Th/Eu]}$ against age in logarithmic scale. The symbols are the same as in Figure \ref{fig:all_res_feh_initial}, with color codes indicating metallicities. We find a significant trend for $\rm{[Th/Eu]}$, but not for $\rm{[Eu/H]}$ and $\rm{[Th/H]}$ (see text).}\label{fig:all_res_age_initial}
\end{figure*}

The presence of the three groups is also visible in the $[\rm{X/H}]$ - $[\rm{Fe/H}]$ and $[\rm{X/H}]$ - age trends in Figure \ref{fig:all_res_feh_initial} and \ref{fig:all_res_age_initial}.
Examining the abundance trends with age in terms of $\rm{[X/Fe]}$ may complicate the discussion because of the age-metallicity relation degeneracy.
Thus we examine the abundance trends in terms of $\rm{[X/H]}$ so as not to complicate the discussion of Eu and Th production with the variation of metallicity with age. 

We fit a linear regression model to our data using the form:
\begin{equation}
    y = m (x - x_{\rm{mean}}) + b
\end{equation}
where $m$ is the slope and $b$ represents the predicted value of $y$ at $x = x_{\rm{mean}}$. 
This form ensures that the intercept is well-constrained without extrapolating beyond the data range while maintaining the robustness of the slope.
We test the significance of the slope by the $p$-value, which indicates the probability of observing a trend when the null hypothesis of zero slope is true. 
Conservatively, a threshold of $p = 0.05$ is assumed to determine the significance of the trend. 
We show the $p$-value, slope, and the uncertainty of the slope for each of our linear regression fittings in Table \ref{tab:linear_fit}.

Our result shows that $\rm{[Eu/H]}$ clearly correlates with metallicity, as shown in the left panel of Figure \ref{fig:all_res_feh_initial}.
The slope of $\rm{[Eu/H]}$ with $\rm{[Fe/H]}$ trend is significant compared to its uncertainty.
The value of $p = 0.0$ confirms this result, thus we conclude that there is a significant correlation between $\rm{[Eu/H]}$ and $\rm{[Fe/H]}$.

On the contrary, we find that the correlation between $\rm{[Eu/H]}$ and age is less clear, close to a flat trend as shown on the left panel of Figure \ref{fig:all_res_age_initial}. 
The linear fit shows a decreasing trend with increasing age, though the slope is not significant compared to the uncertainty.
Indeed, we find a rather high $p$-value for the trend to be significant.
Thus our results suggest that $\rm{[Eu/H]}$ depends on metallicities rather than age.
As found in the left panel of Figure \ref{fig:all_res_age_initial}, old metal-poor stars have lower $\rm{[Eu/H]}$ than old metal-rich stars. 
Among them, four stars particularly have low values.  

\begin{table*}
\centering
\begin{tabular}{lllllll}
\hline
       & $\rm{[Eu/H]}$ - $\rm{[Fe/H]}$ & $\rm{[Th/H]}$ - $\rm{[Fe/H]}$ & $\rm{[Th/Eu]}$ - $\rm{[Fe/H]}$ & $\rm{[Eu/H]}$ - age & $\rm{[Th/H]}$ - age & $\rm{[Th/Eu]}$ - age\\
\hline\hline
$p$      & 0.0  & 0.0  & 0.53 & 0.08 & 0.16 & 0.01   \\
 slope $\pm \sigma_{\rm{slope}}$  & 0.45 $\pm$ 0.04 & 0.49 $\pm$ 0.09 & 0.05 $\pm$ 0.08 & -0.05 $\pm$ 0.03 & 0.07 $\pm$ 0.05 & 0.10 $\pm$ 0.04 \\ 
\hline
\end{tabular}
\caption{Linear regression results for $\rm{[Eu/H]}$, $\rm{[Th/H]}_{\rm{initial}}$, and $\rm{[Th/Eu]}_{\rm{initial}}$ with metallicities and age.}
\label{tab:linear_fit}
\end{table*}

We identify the 4 old stars with the lowest $\rm{[Eu/H]}$ as KIC5737655 ($\rm{[Eu/H]} = -0.25$ at $\rm{[Fe/H]} = -0.63$), 
KIC6531928 ($\rm{[Eu/H]} = -0.25$ at $\rm{[Fe/H]} = -0.57$), KIC10382615 ($\rm{[Eu/H]} = -0.35$ at $\rm{[Fe/H]} = -0.49$), and KIC7734065 ($\rm{[Eu/H]} = -0.36$ at $\rm{[Fe/H]} = -0.43$).
These stars have clear Eu line features, and the uncertainties indicate that our Eu measurement for these stars is robust. 

We also examine the Th features of these stars. 
For KIC5737655 and KIC7734065, we succesfully derive Th abundance and obtain $\rm{[Th/H]_{initial}} = 0.20$ and 0.35, respectively. 
KIC10382615 has a relatively large uncertainty and low $\rm{[Th/Fe]}$ deviating from the main trend in Figure \ref{fig:all_res_feh}, as discussed in Section \ref{section:results}.
We find $\rm{[Th/H]_{initial}} = -0.47$ for this star.
We cannot derive Th for KIC6531928, and instead estimate an upper limit of $\rm{[Th/H]_{initial}} = 0.13$. 
As found in Figure \ref{fig:all_res_age_initial}, $\rm{[Th/H]}_{\rm{initial}}$ of KIC5737655 and KIC7734065 do not deviate from the main trend with age, even though they show a deviation from the trend in $\rm{[Eu/H]}$.
The upper limit of Th for KIC6531928 also does not deviate from the trend, although we cannot make a definite judgment in this case. 
On the other hand, KIC10382615 shows both Eu and Th deviating from the trend, although the large uncertainty of Th for this star hinders us from making a clear conclusion on this matter. 
Interestingly, KIC5737655, KIC7734065, and KIC6531928 are classified as thick disk stars by \citet{Liu2019}, representing three out of five thick disk stars in the sample, while KIC10382615 is identified as a thin disk star.
As these stars do not strongly deviate from the $\rm{[Eu/H]}$ - $\rm{[Fe/H]}$ correlation, their Eu abundance is consistent with the expectations for their metallicity range, particularly at the metal-poor end of the trend.

\begin{figure}[h!]
    \centering
    \includegraphics[width=0.45\textwidth]{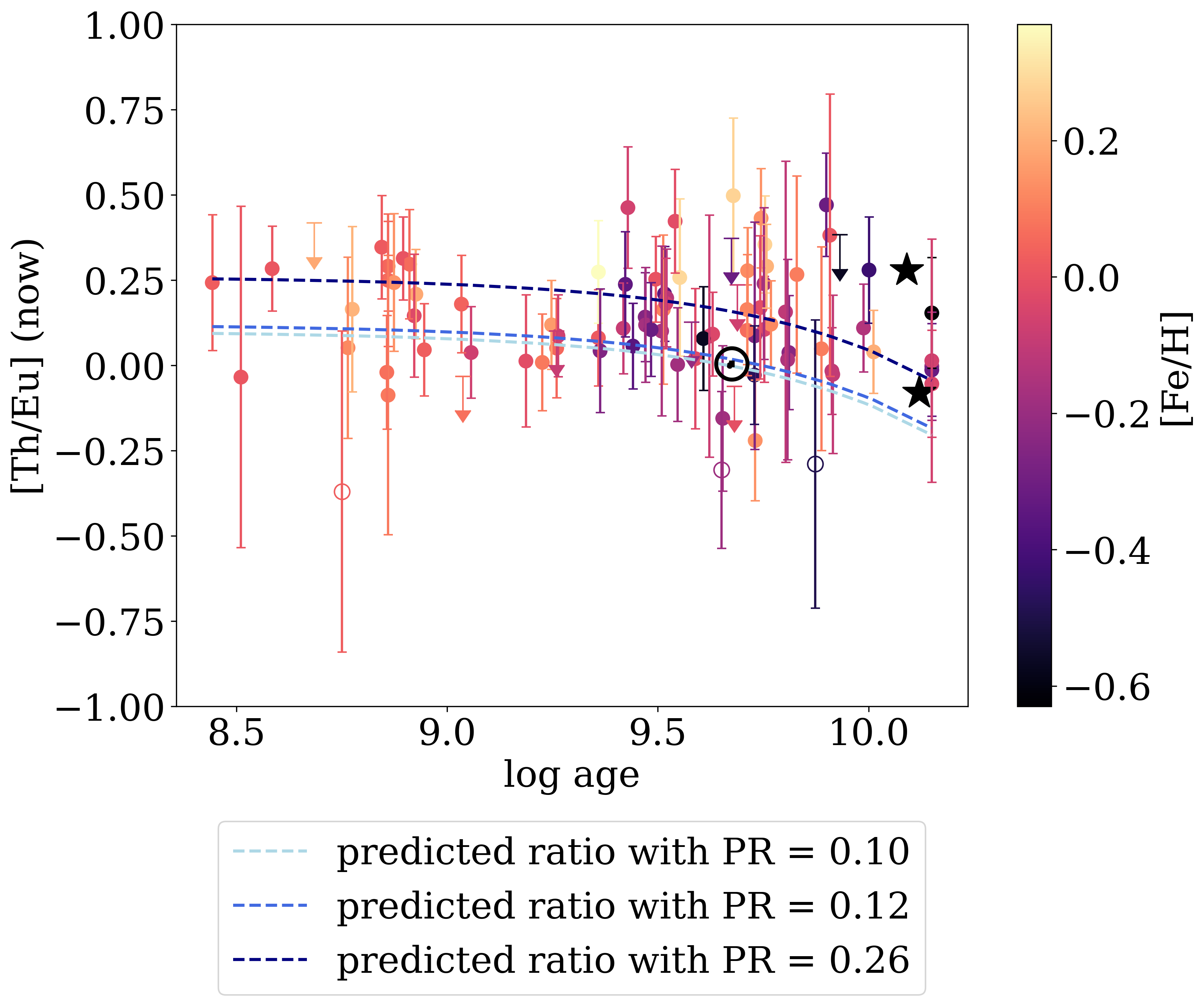}
    
    \caption{Trend of $\rm{[Th/Eu]}_{now}$ with age. The dashed lines show the predicted present-day ratio assuming a constant PR. The solar symbol marks the solar value, and the black stars show the two RPE stars, CS 31082-001 and HE 1523-0901.}\label{fig:age_uncorr}
\end{figure}

The $\rm{[Th/H]}_{\rm{initial}}$ trends are shown in the middle panel of Figure \ref{fig:all_res_feh_initial} and \ref{fig:all_res_age_initial}.
The correction in abundance is negligible for the youngest stars, while an increase of $\rm{[Th/H]}$ (and $\rm{[Th/Eu]}$) reached $\sim 0.3$ dex in the oldest stars. 
The impact of corrections is observable by comparing $\rm{[Th/Eu]}_{\rm{now}}$ and $\rm{[Th/Eu]}_{\rm{initial}}$ in Figure \ref{fig:age_uncorr} and the right panel of Figure \ref{fig:all_res_age_initial}.
In Figure \ref{fig:age_uncorr}, the oldest stars show quite low $\rm{[Th/Eu]}_{\rm{now}}$ compared to the young stars but show higher $\rm{[Th/Eu]}_{\rm{initial}}$ after correction for the decay.

We exclude upper limits and the three stars with the lowest $\rm{[Th/H]}$, which have large uncertainties discussed in Section \ref{section:results}, from our linear regression fittings.
To ensure the robustness of our analysis, we also examine the trends excluding the stars with total Th abundance uncertainties larger than 0.3 dex. 
The slopes of the $\rm{[Th/H]}_{\rm{initial}}$ and $\rm{[Th/Eu]}_{\rm{initial}}$ trends with ages and metallicities exhibit little to no change.
This does not change the significant correlations described below.

We find a significant correlation between $\rm{[Th/H]}_{\rm{initial}}$ and metallicity, as shown in the middle panel of Figure \ref{fig:all_res_feh_initial}.
On the other hand, no significant correlation between $\rm{[Th/H]}_{\rm{initial}}$ and age is found in the middle panel of Figure \ref{fig:all_res_age_initial}.
This suggests that $\rm{[Th/H]}_{\rm{initial}}$, similarly to $\rm{[Eu/H]}$, depends on metallicities rather than ages.
 
No significant correlation is found between $\rm{[Th/Eu]}_{\rm{initial}}$ and metallicity, as shown in the right panel of Figure \ref{fig:all_res_feh_initial}.
In Section \ref{section:results}, we point out that $\rm{[Th/Eu]}_{\rm{now}}$ exhibits a large scatter in the metal-rich region ($\rm{[Fe/H]} \geq -0.1$), where the scatter starts to appear, also in line with the metal-rich group defined in Figure \ref{fig:age_metal}. 
This remains unchanged after correction for decay.
The standard deviation is $\sigma_{\rm{obs}} = 0.15$ for $\rm{[Th/Eu]}_{\rm{now}}$ and $\sigma_{\rm{obs}} = 0.16$ for $\rm{[Th/Eu]}_{\rm{initial}}$.
The overlap of young and old stars in this region suggests that $\rm{[Th/Eu]}$ PR is not constant among stars of different ages.

This is confirmed by the statistically significant correlation between $\rm{[Th/Eu]}_{\rm{initial}}$ and age, as shown in the right panel of Figure \ref{fig:all_res_age_initial}, with a slope of 0.10 $\pm$ 0.04. 
We further confirm the varying PR between the young and old stars by examining the mean $\rm{[Th/Eu]}_{\rm{initial}}$ in each group indicated in Figure \ref{fig:age_metal}.
The mean $\rm{[Th/Eu]}_{\rm{initial}}$ is 0.16 dex for young metal-rich, 0.23 dex for old metal-poor, and 0.29 dex for old metal-rich stars.
The standard error of the mean is 0.03 for each of the three groups, indicating that the difference of the mean $\rm{[Th/Eu]}_{\rm{initial}}$ between the young and either of the old groups is significant.

To further test the evolution of $\rm{[Th/Eu]}_{initial}$ with age, the predicted $\rm{[Th/Eu]}_{now}$ assuming a constant PR is shown in Figure \ref{fig:age_uncorr}.
The Th/Eu PR depends on theoretical calculations, with values ranging from $\log{\rm{(Th/Eu)}} = -0.375$ to $-0.24$ \citep{Cowan1999, Kratz2007, Hill2017, Placco2017, Ji2018}. 
This corresponds to $\rm{[Th/Eu]}_{initial}$ ranging from 0.12 to 0.26 dex. 
In addition, if the scaled solar value is assumed to be the universal PR, it would give $\rm{[Th/Eu]}_{initial} = 0.10$ dex given the age of the solar system, $\sim$4.6 Gyr ($\log{\rm{age}} = 9.7$, the present solar value is shown by the black solar symbol). 
Figure \ref{fig:age_uncorr} shows that a constant PR, regardless of the assumed value, cannot account for the $\rm{[Th/Eu]}_{now}$ variation in old stars. 
This shows that Th decay alone cannot explain the observed spread in $\rm{[Th/Eu]}_{now}$.

Our findings suggest that while $\rm{[Th/Eu]}$ PR varies over time, it does not depend on metallicity. 
This aligns with our findings, as both Eu and Th are expected to increase with metallicity. 
Further interpretation of these findings will be discussed in the following section.

\subsection{The implications on the r-process production}
In this section, we discuss the interpretation of the correlations described above. 
First, we observe positive correlations for both Eu and Th with metallicities.
The difference in slope of $\rm{[Eu/H]}$ - $\rm{[Fe/H]}$ and $\rm{[Th/H]}_{\rm{initial}}$ - $\rm{[Fe/H]}$ is 0.04 $\pm$ 0.10, showing that there is no significant difference in the strength of those correlations.
This is further supported by the flat trend of $\rm{[Th/Eu]}_{\rm{initial}}$ against metallicity (see also \citet{Roederer2009}).
This finding aligns with standard GCE models, where the production of Eu and Th increases as the events that produce Fe also increase.

Second, we find a statistically significant positive correlation between $\rm{[Th/Eu]}_{initial}$ and age.
This may hint at a more efficient Th production in the early universe, causing a slight increase of $\rm{[Th/Eu]}_{initial}$ ratio towards older age.
Such a scenario can be explained by different dominant r-process source(s) in the early and present-day universe, i.e., short and long-timescale events.
As the universe evolves, the relative contributions of different sites of r-process events change, leading to the $\rm{[Th/Eu]}_{initial}$ ratio's dependency on age. There have been discussions supporting the combination of long and short-timescale r-process sources, e.g., \citet{Cote2019}, \citet{Mishenina2022}, and \citet{Farouqi2022}. However, there is no clear consensus on short-timescale source(s) that can produce a high actinide-to-lanthanide ratio. 

Even though CCSNe are commonly proposed as a candidate for a short timescale r-process source, the physical conditions do not favor the heavy r-process element production \citep{Wanajo2011, Janka2012, Goriely2016}. 
Instead, compact binary mergers such as NSMs or BH-NS mergers are more favored, as their ejecta are more neutron-rich, resulting in lanthanides and actinides production \citep{Wanajo2014, Just2015}.
Rare types of CCNe, such as MRD SNe, may produce actinides, but likely only in small quantities, indicating that this type of event contributes a minimal fraction of the universal r-process material \citep{Reichert2023}. 
However, the current models for CCSNe and NSMs still have limitations in their detailed treatment of the underlying physics (see, for example, \citet{Goriely2016}). 
Therefore, further investigation is necessary to develop a model that can produce actinides on a short timescale under realistic astrophysical conditions.

It is also crucial to contextualize our results within the universal $\rm{[Th/Eu]}$ ratios, including the more metal-poor regime not covered in our sample.
Metal-poor stars exhibit a large scatter of $> 1.0$ dex in $\rm{[Th/Eu]}_{now}$ at $\rm{[Fe/H]} \leq -1.8$ (see, for example, \citet{Ren2012} and Figure 5 of \citet{Wu2022}).
Since these stars are very metal-poor, it is reasonable to assume they have ages around 13 Gyr, which would lead to a similar amount of scatter in $\rm{[Th/Eu]}_{initial}$.
The abundance scatter in metal-poor stars is commonly explained by the environmental factors of their birthplace.
These stars typically reside in the galactic halo and smaller systems, where stochastic r-process events in the early universe and inhomogeneous mixing of the natal cloud naturally result in a large star-to-star abundance scatter. 

In Figure \ref{fig:age_uncorr}, we present two examples of RPE halo stars, CS 31082-001 and HE 1523-0901 (black star symbols), previously analyzed by \citet{Cayrel2001} and \citet{Frebel2007}. 
Their ages are estimated by employing various chronometers, including U/Th, U/Ir, and U/Os, resulting in a median age of 12.5 Gyr for CS 31082-001 \citep{Cayrel2001} and a weighted average of 13.2 Gyr for HE 1523-0901 \citep{Frebel2007}.
The current $\rm{[Th/Eu]}_{now}$ value for CS 31082-001 exceeds the predicted ratio from decay, even when assuming the highest PR, while the value for HE 1523-0901 falls between the highest and lowest PRs.
Although this may support the variation of PR found in our study, it cannot be directly concluded for these stars due to the environmental factors mentioned above.
On the other hand, in our sample of disk stars, a larger number of events and a more homogeneous chemical mixing are expected to contribute to the observed stellar abundances.
Even though some degree of inhomogeneous mixing in the disk cannot be ruled out, especially in the case of old stars, the wide age distribution of our sample allows for a more robust analysis of PR variation.
Therefore, our study shows that while $\rm{[Th/Eu]}_{initial}$ is correlated with age, environmental factors also play an important role in shaping the $\rm{[Th/Eu]}$ distribution.

\subsection{Further improvements and future prospects}
The improved precision of the Th abundance will provide a more stringent constraint on the r-process evolution, especially the actinides, in the metal-rich regime.
To obtain more accurate and reliable abundances, an improvement could be made to the measurements of the atomic properties of this line. 
As mentioned in Section \ref{section:methods}, there is quite a large uncertainty, 10\% of the $gf$ value, in the Th line analyzed in this study.
Improving the precision of the oscillator strength would minimize the systematic offset in the Th abundance measurement.

Another important point is compiling a complete line list including all the atomic and molecular lines in the region, especially the strong lines most likely to appear in the stellar spectra. 
This is especially crucial for lines that might be a source of blending in the Th line as in the case of the CN 5989.054 \AA\ line.
The precision of the other lines' data is also of importance, for example, the oscillator strengths of the neighboring Si I lines and the wavelength of the Nd II line.
While these lines do not directly overlap with the Th line, the overall fitting accuracy would improve the local continuum placement.

Furthermore, as the uncertainties in the Th abundance measurement are dominated by the fitting uncertainty, increasing the SNR in the Th region will help reduce the measurement uncertainties.
This is especially important in cases where the Th line feature is weak.

Analyzing a more homogeneous sample such as globular clusters would help to better constrain the abundance trends with metallicities.
As globular clusters consist of stars with similar ages, there would be no overlap between different populations of stars, minimizing the noise in analyzing the abundance trends. 
This would provide a constraint for old metal-poor stars around $-2.5 \leq \rm{[Fe/H]} \leq -0.5$.
The constraint for young metal-rich stars could be provided by studying open clusters in the metallicity range of $-0.5 \leq \rm{[Fe/H]} \leq 0.3$.
Combining both types of samples would then give a complete picture of the abundance trends with age, without overlapping the populations.
Such studies are currently scarce due to the difficulty in detecting the Th line, some examples being \citet{Sneden2000a} and \citet{Yong2008}.
However, the small sample sizes hinder any conclusion on the Th abundance trend.

The correlation of the $\rm{[Th/Eu]}_{initial}$ ratio with stellar age indicates that the Th/Eu PR is not constant. 
This emphasizes the need to constrain its evolution through both observational data and theoretical calculations. 
Previous discussions of the varying PR can be found in works by \citet{Hill2017} and \citet{Ji2018}, which demonstrate that in some cases, tuning the PR is necessary to derive a physical stellar age using the Th/Eu chronometer. 
This suggests that using the Th/Eu ratio as a chronometer should be done with caution.
An alternative is the Th/U chronometer, which may offer reduced systematic uncertainties in the PR due to the close mass numbers of Th and U \citep{Cowan1999, Schatz2002}. 
However, this approach comes with the challenge of detecting U lines and accurately determining their abundance.

Our results suggest that Th is more efficiently produced in the early universe, pointing to the necessity of actinide production in short-timescale source(s).
Currently, there is insufficient evidence supporting any particular source, even though NSMs are preferred over CCSNe in terms of physical conditions, despite concerns about delay times.
This study provides observational constraints for future research into theoretical models of actinide production in the early universe.

\section{Conclusions}\label{section:conclusions}
We have derived abundance measurements for Th and Eu for a sample of red giant stars with $-0.7 \leq \rm{[Fe/H]} \leq 0.4$. 
We demonstrate that Th abundance can be derived from the Th II line at 5989 \AA\ line with higher precision, by careful treatment of the atomic data and blending. 
While we find that the $[\rm{Eu/Fe}]$ abundance is well constrained, there is a larger spread on $[\rm{Th/Fe}]$. 
For the first time, we derive the initial Th abundance using prior age information from seismology.
While individual measurement uncertainty may be considerable, our large sample size enables us to examine statistically significant correlations between abundance ratios with metallicities and ages.

We identify the following key findings:
\begin{enumerate}
    \item There is a significant correlation between ${\rm{[Eu/H]}}$ and ${\rm{[Fe/H]}}$, as well as ${\rm{[Th/H]}_{\rm{initial}}}$ and ${\rm{[Fe/H]}}$, as expected from standard GCE models. 
    \item On the other hand, there is no significant trend between ${\rm{[Th/Eu]}_{initial}}$ and ${\rm{[Fe/H]}}$, suggesting that the Th/Eu ratio remains broadly uniform across metallicities covered in our sample. 
    \item The abundance ratio $\rm{[Th/Eu]_{initial}}$ shows a significant correlation with age, which may point to different dominant r-process site contributions in the early universe compared to the present day.
\end{enumerate}

Our analysis can be improved by minimizing the Th abundance uncertainties, e.g., by improving the precision of the line data, completing the line list, and increasing the spectra SNR. 
More accurate Th abundance measurement would benefit the understanding of the actinide evolution in metal-rich stars.

\begin{acknowledgements}
This research is based on data collected at Subaru Telescope, which is operated by the National Astronomical Observatory of Japan. We are honored and grateful for the opportunity of observing the Universe from Maunakea, which has cultural, historical, and natural significance in Hawaii. 
The data are retrieved from the JVO portal (\href{http://jvo.nao.ac.jp/portal}{http://jvo.nao.ac.jp/portal}) operated by NAOJ.
This work was supported in part by JSPS KAKENHI Grant Numbers JP21H04499, JP20H05855.
This was partially supported by the Spinoza grant that was awarded to Amina Helmi.
TM is supported by a Gliese Fellowship at the Zentrum für Astronomie, University of Heidelberg, Germany. 
AA is grateful to Nicholas Storm for helpful discussions regarding the use of TSFitPy.

\end{acknowledgements}

\bibliographystyle{aasjournal}
\bibliography{biblio}

\begin{appendix}
    
\section{Additional Tables}

\begin{table}[ht]
\centering
\caption{Line list for the Th 5989 \AA\ region. The lines shown in this table include the atomic lines from the GES line list and the CN lines from \citet{Brooke2014}. The lines of ${}^{}\rm{TiO}$ and ${}^{12}\rm{C}{}^{12}\rm{C}$, which we also take into account in our calculations, are not shown here, but available on the online database as described in Section \ref{section:methods}. The lines marked with an asterisk are the ones for which we made adjustments as described in Section \ref{section:methods}.}\label{tab:linelist}

\begin{tabular}{llll}
\hline
Wavelength & $\chi$ & $\log{gf}$ & Element   \\
\hline
\hline
5988.791   & 5.964          & -2.053  & Si I *\\
5988.838   & 5.964          & -9.99 & Si I *\\
5988.855   & 0.751          & -3.52  & ${}^{12}\rm{C}{}^{14}\rm{N}$       \\
5988.875   & 2.095          & -9.76  & ${}^{12}\rm{C}{}^{14}\rm{N}$        \\
5988.898   & 2.387          & -3.21  & ${}^{12}\rm{C}{}^{14}\rm{N}$       \\
5988.899   & 2.949          & -2.380 & Ce II    \\
5988.93    & 0.661          & -3.07    & ${}^{13}\rm{C}{}^{14}\rm{N}$     \\
5988.955   & 0.262          & -13.16 & V I     \\
5988.958   & 12.240         & -5.731 & Cr II    \\
5988.987   & 2.347          & -2.87  & ${}^{13}\rm{C}{}^{14}\rm{N}$   \\
5988.994   & 8.097          & -0.091 & Ti II    \\
5989.0     & 2.347          & -2.85    & ${}^{13}\rm{C}{}^{14}\rm{N}$   \\
5989.009   & 4.208          & -4.728 & Cr I     \\
5989.027   & 12.240         & -5.214 & Cr II    \\
5989.03    & 7.946          & -3.163   & C I     \\
5989.033   & 3.399          & -2.754 & Ti I     \\
5989.045   & 11.092         & -9.474 & Mn II    \\
5989.045   & 0.189          & -1.414 & Th II    \\
5989.054   & 0.897          & -2.23  & ${}^{12}\rm{C}{}^{14}\rm{N}$ *  \\
5989.083   & 3.042          & -3.68  & ${}^{12}\rm{C}{}^{14}\rm{N}$ \\
5989.106   & 2.474          & -2.23  & ${}^{12}\rm{C}{}^{14}\rm{N}$ \\
5989.158   & 1.644          & -2.090 & Ce II    \\
5989.181   & 6.145          & -3.109 & Cr II    \\
5989.197   & 1.506          & -3.11  & ${}^{13}\rm{C}{}^{14}\rm{N}$  \\
5989.235   & 6.671          & -3.868 & Mn II    \\
5989.24    & 3.048          & -3.02    & ${}^{12}\rm{C}{}^{14}\rm{N}$ \\
5989.249   & 2.339          & -3.40  & ${}^{13}\rm{C}{}^{14}\rm{N}$   \\
5989.252   & 2.339          & -3.70  & ${}^{13}\rm{C}{}^{14}\rm{N}$   \\
5989.256   & 14.912         & -4.620 & Ni II    \\
5989.261   & 2.450          & -2.08  & ${}^{12}\rm{C}{}^{14}\rm{N}$  \\
5989.282   & 14.903         & -5.590 & Ni II    \\
5989.287   & 3.048          & -3.09  & ${}^{12}\rm{C}{}^{14}\rm{N}$  \\
5989.289   & 6.285          & -3.482 & Cr II    \\
5989.294   & 2.450          & -3.04  & ${}^{12}\rm{C}{}^{14}\rm{N}$ \\
5989.3     & 1.865          & -5.16   & ${}^{13}\rm{C}{}^{14}\rm{N}$   \\
5989.312   & 0.380          & -2.050 & Nd II *\\

\hline
\end{tabular}
\end{table}

\begin{table}[ht]
\centering
\caption{Wavelengths, excitation potentials, oscillator strengths, and the lower and upper energy levels of the line transition of the CN lines described in Section \ref{section:methods}. The data are taken from the line list compiled by \citet{Sneden2014} based on the measurements by \citet{Brooke2014}, and from the online line list of Kurucz for the line at 5989 \AA.}\label{tab:CN_lines}
\begin{tabular}{llllll}
\hline
Wavelength (\AA) & $\chi$ (eV) & $\log{gf}$ & $J_{\rm{low}}$ & $J_{\rm{up}}$ & CN isotopes  \\
\hline
\hline
5972.985   & 0.760  & -2.05   & 32.5   & 32.5  & ${}^{12}\rm{C}{}^{14}\rm{N}$ \\
5989.054   & 0.897  & -2.23   & 41.5   & 42.5  & ${}^{12}\rm{C}{}^{14}\rm{N}$ \\
6644.304   & 0.805  & -2.25   & 14.5   & 15.5  & ${}^{12}\rm{C}{}^{14}\rm{N}$ \\
6644.352   & 0.805  & -2.66   & 15.5   & 15.5  & ${}^{12}\rm{C}{}^{14}\rm{N}$ \\
6644.922   & 1.066  & -1.97   & 59.5   & 60.5  & ${}^{12}\rm{C}{}^{14}\rm{N}$\\
6644.965 & 1.664 & -1.95 & 31.5 & 30.5 & ${}^{13}\rm{C}{}^{14}\rm{N}$ \\
\hline
\end{tabular}
\end{table}

\onecolumn

\longtab[3]{
\setlength{\tabcolsep}{3.5pt}
\begin{longtable}{llllllllllllll}
\caption{Abundance results for Arcturus and our main sample. Stellar parameters are adopted from \citet{Ramirez2011} for Arcturus, and from \citet{Takeda2015} and \citet{Takeda+(2016)} for the main sample. Stars without displayed Th and $\rm{[Th/Eu]}$ uncertainties are the ones for which we derived upper limits.}\label{tab:redgiants}
\hline
ID  & $\log{\rm{(age)}}$ & $\rm{[Fe/H]}$  & $T_{\rm{eff}}$ (K)  & $\log{g}$ & $v_{\rm{mic}}$ & $\rm{[Eu/H]}$  & $\rm{[Eu/Fe]}$ & $\sigma_{\rm{Eu}}$ & $\rm{[Th/H]}$  & $\rm{[Th/Fe]}$ & $\sigma_{\rm{Th}}$ & ${\rm{[Th/Eu]}}$ & $\sigma_{\rm{Th/Eu}}$\\
\hline
\hline
\endfirsthead
\caption{continued.}\\
\hline
ID  & $\log{\rm{(age)}}$ & $\rm{[Fe/H]}$  & $T_{\rm{eff}}$ (K)   & $\log{g}$ & $v_{\rm{mic}}$ & $\rm{[Eu/H]}$  & $\rm{[Eu/Fe]}$ & $\sigma_{\rm{Eu}}$ & $\rm{[Th/H]}$  & $\rm{[Th/Fe]}$ & $\sigma_{\rm{Th}}$ & ${\rm{[Th/Eu]}}$ & $\sigma_{\rm{Th/Eu}}$ \\
\hline
\hline
\endhead

Arcturus    &    & -0.52 & 4286 & 1.66 & 1.74 & -0.22 & 0.30 & 0.04 & -0.34 & 0.18 & 0.05 & -0.12 & 0.06\\
KIC4770846  & 9.359    & 0.02  & 4847 & 2.6  & 1.27   & 0.06  & 0.04        & 0.067          & 0.14  & 0.12        & 0.12           & 0.08  & 0.14                \\
KIC5000307  & 9.47     & -0.25 & 5023 & 2.64 & 1.27   & -0.06 & 0.19        & 0.067          & 0.08  & 0.33        & 0.11           & 0.14  & 0.13                \\
KIC5266416  & 9.418    & -0.09 & 4767 & 2.5  & 1.33   & 0.02  & 0.11        & 0.066          & 0.13  & 0.22        & 0.12           & 0.11  & 0.13                \\
KIC5737655  & 10.15    & -0.63 & 5026 & 2.45 & 1.46   & -0.25 & 0.38        & 0.068          & -0.1  & 0.53        & 0.15           & 0.15  & 0.16                \\
KIC11819760 & 9.654    & -0.18 & 4824 & 2.36 & 1.32   & -0.01 & 0.17        & 0.066          & -0.16 & 0.02        & 0.2            & -0.15 & 0.21                \\
KIC2573092  & 9.495    & 0.0   & 4689 & 2.48 & 1.38   & 0.02  & 0.02        & 0.07           & 0.27  & 0.27        & 0.1            & 0.25  & 0.13                \\
KIC3098045  & 9.811    & -0.24 & 4820 & 2.34 & 1.29   & -0.16 & 0.08        & 0.068          & -0.12 & 0.12        & 0.15           & 0.04  & 0.17                \\
KIC3323943  & 9.988    & -0.14 & 4826 & 2.55 & 1.26   & 0.04  & 0.18        & 0.069          & 0.15  & 0.29        & 0.11           & 0.11  & 0.13                \\
KIC3425476  & 9.589    & -0.03 & 4780 & 2.57 & 1.3    & 0.03  & 0.06        & 0.068          & 0.05  & 0.08        & 0.19           & 0.02  & 0.21                \\
KIC4039306  & 9.915    & -0.1  & 4806 & 2.45 & 1.3    & -0.04 & 0.06        & 0.067          & -0.07 & 0.03        & 0.22           & -0.03 & 0.23                \\
KIC5514974  & 9.913    & -0.1  & 4674 & 2.26 & 1.3    & -0.03 & 0.07        & 0.066          & -0.05 & 0.05        & 0.11           & -0.02 & 0.13                \\
KIC5858034  & 9.651    & -0.19 & 4887 & 2.45 & 1.28   & -0.13 & 0.06        & 0.067          & -0.44 & -0.25       & 0.3            & -0.31 & 0.3                 \\
KIC7734065  & 10.001   & -0.43 & 4882 & 2.2  & 1.28   & -0.36 & 0.07        & 0.07           & -0.08 & 0.35        & 0.14           & 0.28  & 0.16                \\
KIC10382615 & 9.873    & -0.49 & 4890 & 2.25 & 1.27   & -0.35 & 0.14        & 0.07           & -0.64 & -0.15       & 0.49           & -0.29 & 0.5                 \\
KIC10600926 & 10.15    & -0.2  & 4879 & 2.48 & 1.29   & 0.02  & 0.22        & 0.068          & 0.02  & 0.22        & 0.14           & 0.0   & 0.16                \\
KIC10604460 & 9.731    & 0.14  & 4573 & 2.37 & 1.3    & 0.21  & 0.07        & 0.069          & -0.01 & -0.15       & 0.16           & -0.22 & 0.18                \\
KIC11177749 & 9.753    & -0.04 & 4677 & 2.24 & 1.26   & -0.08 & -0.04       & 0.067          & 0.03  & 0.07        & 0.14           & 0.11  & 0.16                \\
KIC11352756 & 10.15    & -0.04 & 4604 & 2.29 & 1.31   & 0.1   & 0.14        & 0.066          & 0.05  & 0.09        & 0.14           & -0.05 & 0.16                \\
KIC12008680 & 10.15    & -0.32 & 4881 & 2.55 & 1.31   & 0.07  & 0.39        & 0.066          & 0.06  & 0.38        & 0.12           & -0.01 & 0.14                \\
KIC2013502  & 9.187    & -0.02 & 4913 & 2.69 & 1.14   & 0.02  & 0.04        & 0.07           & 0.03  & 0.05        & 0.18           & 0.01  & 0.19                \\
KIC2448225  & 9.248    & 0.16  & 4577 & 2.37 & 1.32   & 0.07  & -0.09       & 0.069          & 0.19  & 0.03        & 0.11           & 0.12  & 0.13                \\
KIC3730953  & 9.057    & -0.07 & 4861 & 2.55 & 0.97   & -0.02 & 0.05        & 0.066          & 0.02  & 0.09        & 0.12           & 0.04  & 0.13                \\
KIC3758458  & 9.038    & 0.07  & 5009 & 2.71 & 1.28   & 0.1   & 0.03        & 0.066          & 0.07  & 0.0         & -              & -0.03 & -                   \\
KIC4902641  & 8.751    & 0.03  & 4987 & 2.84 & 1.15   & 0.08  & 0.05        & 0.067          & -0.29 & -0.32       & 0.54           & -0.37 & 0.54                \\
KIC5088362  & 8.946    & 0.03  & 4760 & 2.41 & 1.33   & 0.02  & -0.01       & 0.067          & 0.07  & 0.04        & 0.12           & 0.05  & 0.14                \\
KIC5128171  & 9.034    & 0.04  & 4808 & 2.54 & 1.3    & 0.04  & -0.0        & 0.068          & 0.22  & 0.18        & 0.13           & 0.18  & 0.14                \\
KIC5307747  & 8.585    & 0.01  & 5031 & 2.77 & 1.27   & 0.04  & 0.03        & 0.066          & 0.32  & 0.31        & 0.1            & 0.28  & 0.12                \\
KIC5990753  & 8.685    & 0.19  & 5011 & 2.97 & 1.15   & 0.17  & -0.02       & 0.068          & 0.59  & 0.4         & -              & 0.42  & -                   \\
KIC6276948  & 8.926    & 0.19  & 4939 & 2.84 & 1.18   & 0.19  & 0.0         & 0.068          & 0.4   & 0.21        & 0.11           & 0.21  & 0.13                \\
KIC7205067  & 8.443    & 0.03  & 5064 & 2.58 & 1.32   & -0.03 & -0.06       & 0.067          & 0.21  & 0.18        & 0.19           & 0.24  & 0.2                 \\
KIC7581399  & 8.511    & 0.01  & 5070 & 2.74 & 1.14   & 0.03  & 0.02        & 0.068          & 0.0   & -0.01       & 0.5            & -0.03 & 0.5                 \\
KIC8378462  & 8.86     & 0.06  & 4995 & 2.81 & 1.19   & 0.1   & 0.04        & 0.069          & 0.35  & 0.29        & 0.18           & 0.25  & 0.19                \\
KIC9173371  & 8.896    & 0.0   & 5064 & 2.85 & 1.18   & 0.04  & 0.04        & 0.066          & 0.35  & 0.35        & 0.1            & 0.31  & 0.12                \\
KIC2988638  & 8.911    & 0.06  & 4912 & 2.67 & 1.22   & 0.0   & -0.06       & 0.067          & 0.3   & 0.24        & 0.15           & 0.3   & 0.16                \\
KIC4056266  & 8.922    & -0.03 & 5021 & 2.67 & 1.17   & 0.01  & 0.04        & 0.069          & 0.16  & 0.19        & 0.17           & 0.15  & 0.18                \\
KIC4570120  & 8.86     & 0.05  & 5035 & 2.73 & 1.2    & 0.03  & -0.02       & 0.069          & 0.32  & 0.27        & 0.11           & 0.29  & 0.13                \\
KIC5283798  & 9.226    & 0.09  & 4770 & 2.53 & 1.3    & 0.09  & 0.0         & 0.066          & 0.1   & 0.01        & 0.13           & 0.01  & 0.14                \\
KIC5611192  & 8.845    & 0.02  & 5064 & 2.91 & 1.2    & 0.11  & 0.09        & 0.072          & 0.46  & 0.44        & 0.13           & 0.35  & 0.15                \\
KIC9349632  & 8.765    & 0.12  & 4976 & 2.75 & 1.12   & 0.09  & -0.03       & 0.068          & 0.14  & 0.02        & 0.26           & 0.05  & 0.27                \\
KIC9583430  & 8.775    & 0.21  & 4854 & 2.73 & 1.19   & 0.09  & -0.12       & 0.067          & 0.25  & 0.04        & 0.23           & 0.16  & 0.24                \\
KIC10474071 & 8.86     & 0.09  & 4975 & 2.65 & 1.17   & 0.01  & -0.08       & 0.07           & -0.08 & -0.17       & 0.4            & -0.09 & 0.41                \\
KIC11251115 & 8.857    & 0.07  & 4837 & 2.57 & 1.29   & 0.1   & 0.03        & 0.066          & 0.08  & 0.01        & 0.15           & -0.02 & 0.17                \\
KIC11721438 & 8.874    & 0.14  & 4959 & 2.87 & 1.12   & 0.14  & -0.0        & 0.068          & 0.38  & 0.24        & 0.19           & 0.24  & 0.2                 \\
KIC12070114 & 9.26     & 0.05  & 4698 & 2.46 & 1.3    & 0.05  & -0.0        & 0.069          & 0.1   & 0.05        & 0.13           & 0.05  & 0.15                \\
KIC3455760  & 9.263    & -0.07 & 4654 & 2.68 & 1.13   & 0.15  & 0.22        & 0.067          & 0.24  & 0.31        & 0.1            & 0.09  & 0.12                \\
KIC3744043  & 9.516    & -0.35 & 4946 & 2.94 & 1.1    & -0.14 & 0.21        & 0.067          & 0.07  & 0.42        & 0.12           & 0.21  & 0.14                \\
KIC4243623  & 9.9      & -0.31 & 5005 & 3.75 & 0.74   & 0.09  & 0.4         & 0.071          & 0.56  & 0.87        & 0.13           & 0.47  & 0.15                \\
KIC4351319  & 9.552    & 0.29  & 4876 & 3.32 & 0.91   & 0.15  & -0.14       & 0.073          & 0.41  & 0.12        & 0.22           & 0.26  & 0.23                \\
KIC4952717  & 9.713    & 0.13  & 4793 & 3.13 & 0.93   & 0.18  & 0.05        & 0.068          & 0.46  & 0.33        & 0.11           & 0.28  & 0.13                \\
KIC5033245  & 9.513    & 0.11  & 5049 & 3.41 & 0.97   & 0.06  & -0.05       & 0.068          & 0.22  & 0.11        & 0.21           & 0.16  & 0.22                \\
KIC5530598  & 9.359    & 0.37  & 4599 & 2.85 & 1.04   & 0.25  & -0.12       & 0.079          & 0.52  & 0.15        & 0.13           & 0.27  & 0.15                \\
KIC5723165  & 9.541    & -0.02 & 5255 & 3.67 & 0.9    & 0.01  & 0.03        & 0.07           & 0.43  & 0.45        & 0.13           & 0.42  & 0.15                \\
KIC5806522  & 9.768    & 0.12  & 4574 & 2.6  & 1.09   & 0.18  & 0.06        & 0.067          & 0.3   & 0.18        & 0.11           & 0.12  & 0.13                \\
KIC5866737  & 9.363    & -0.26 & 4874 & 2.86 & 1.14   & -0.04 & 0.22        & 0.067          & 0.0   & 0.26        & 0.17           & 0.04  & 0.18                \\
KIC6117517  & 9.754    & 0.28  & 4649 & 2.94 & 0.97   & 0.21  & -0.07       & 0.07           & 0.56  & 0.28        & 0.12           & 0.35  & 0.14                \\
KIC6144777  & 9.745    & 0.14  & 4734 & 3.02 & 1.02   & 0.17  & 0.03        & 0.07           & 0.6   & 0.46        & 0.13           & 0.43  & 0.15                \\
KIC8702606  & 9.803    & -0.11 & 5472 & 3.66 & 1.01   & -0.07 & 0.04        & 0.074          & 0.09  & 0.2         & 0.44           & 0.16  & 0.44                \\
KIC8718745  & 9.73     & -0.25 & 4902 & 3.2  & 0.98   & 0.08  & 0.33        & 0.069          & 0.17  & 0.42        & 0.33           & 0.09  & 0.33                \\
KIC2696732  & 9.521    & -0.13 & 4821 & 2.9  & 1.04   & 0.01  & 0.14        & 0.067          & 0.21  & 0.34        & 0.13           & 0.2   & 0.14                \\
KIC3531478  & 9.429    & -0.06 & 5000 & 3.2  & 0.98   & -0.05 & 0.01        & 0.071          & 0.41  & 0.47        & 0.16           & 0.46  & 0.18                \\
KIC4350501  & 9.26     & -0.09 & 4929 & 3.19 & 1.01   & 0.11  & 0.2         & 0.07           & 0.21  & 0.3         & -              & 0.1   & -                   \\
KIC4448777  & 9.83     & 0.1   & 4805 & 3.19 & 0.95   & 0.12  & 0.02        & 0.068          & 0.39  & 0.29        & 0.28           & 0.27  & 0.29                \\
KIC4726049  & 9.509    & -0.16 & 5029 & 3.25 & 0.97   & -0.03 & 0.13        & 0.07           & 0.07  & 0.23        & 0.24           & 0.1   & 0.25                \\
KIC5598645  & 9.752    & -0.17 & 5052 & 3.44 & 0.86   & -0.09 & 0.08        & 0.07           & 0.15  & 0.32        & 0.21           & 0.24  & 0.22                \\
KIC6531928  & 9.932    & -0.57 & 5156 & 3.73 & 0.69   & -0.25 & 0.32        & 0.074          & 0.13  & 0.7         & -              & 0.38  & -                   \\
KIC6579495  & 9.743    & -0.02 & 4772 & 2.78 & 1.03   & 0.08  & 0.1         & 0.066          & 0.25  & 0.27        & 0.2            & 0.17  & 0.21                \\
KIC6665058  & 9.689    & -0.07 & 4750 & 3.1  & 0.86   & 0.09  & 0.16        & 0.068          & 0.33  & 0.4         & -              & 0.24  & -                   \\
KIC7584122  & 9.622    & -0.08 & 4974 & 3.29 & 0.93   & 0.06  & 0.14        & 0.07           & 0.15  & 0.23        & 0.35           & 0.09  & 0.36                \\
KIC7799349  & 9.679    & 0.28  & 4969 & 3.56 & 0.94   & 0.08  & -0.2        & 0.072          & 0.58  & 0.3         & 0.22           & 0.5   & 0.23                \\
KIC8475025  & 9.519    & -0.06 & 4848 & 2.88 & 1.02   & 0.06  & 0.12        & 0.069          & 0.24  & 0.3         & 0.11           & 0.18  & 0.13                \\
KIC8493735  & 9.908    & 0.01  & 5842 & 3.62 & 1.14   & 0.05  & 0.04        & 0.156          & 0.43  & 0.42        & 0.38           & 0.38  & 0.41                \\
KIC8751420  & 9.808    & -0.15 & 5260 & 3.63 & 0.95   & 0.07  & 0.22        & 0.068          & 0.09  & 0.24        & 0.29           & 0.02  & 0.29                \\
KIC9145955  & 9.423    & -0.34 & 4943 & 2.85 & 1.05   & -0.11 & 0.23        & 0.068          & 0.13  & 0.47        & 0.14           & 0.24  & 0.15                \\
KIC9812421  & 9.58     & -0.21 & 5140 & 3.48 & 0.87   & -0.04 & 0.17        & 0.068          & 0.09  & 0.3         & -              & 0.13  & -                   \\
KIC10709799 & 9.63     & -0.04 & 4522 & 2.51 & 1.11   & 0.2   & 0.24        & 0.067          & 0.29  & 0.33        & 0.1            & 0.09  & 0.12                \\
KIC10866415 & 9.682    & -0.01 & 4791 & 2.82 & 0.93   & 0.05  & 0.06        & 0.071          & -0.01 & 0.0         & -              & -0.06 & -                   \\
KIC11401156 & 9.888    & 0.1   & 5053 & 3.58 & 0.9    & 0.13  & 0.03        & 0.069          & 0.18  & 0.08        & 0.29           & 0.05  & 0.3                 \\
KIC11618103 & 9.471    & -0.17 & 4922 & 2.91 & 1.1    & -0.09 & 0.08        & 0.066          & 0.03  & 0.2         & 0.15           & 0.12  & 0.17                \\
KIC11717120 & 9.675    & -0.31 & 5087 & 3.72 & 0.75   & 0.02  & 0.33        & 0.068          & 0.39  & 0.7         & -              & 0.37  & -                   \\
KIC11802968 & 10.15    & -0.06 & 4962 & 3.73 & 0.75   & 0.18  & 0.24        & 0.067          & 0.19  & 0.25        & 0.35           & 0.01  & 0.36                \\
KIC1726211  & 9.608    & -0.57 & 4983 & 2.49 & 1.34   & -0.14 & 0.43        & 0.067          & -0.06 & 0.51        & 0.14           & 0.08  & 0.15                \\
KIC2303367  & 9.712    & 0.06  & 4601 & 2.39 & 1.36   & 0.1   & 0.04        & 0.066          & 0.2   & 0.14        & 0.12           & 0.1   & 0.13                \\
KIC2424934  & 9.547    & -0.18 & 4792 & 2.48 & 1.31   & -0.04 & 0.14        & 0.066          & -0.04 & 0.14        & 0.15           & 0.0   & 0.17                \\
KIC2714397  & 9.729    & -0.47 & 4910 & 2.56 & 1.38   & -0.11 & 0.36        & 0.069          & -0.14 & 0.33        & 0.13           & -0.03 & 0.14                \\
KIC3217051  & 9.758    & 0.21  & 4590 & 2.44 & 1.27   & 0.12  & -0.09       & 0.07           & 0.41  & 0.2         & 0.1            & 0.29  & 0.12                \\
KIC4036007  & 9.441    & -0.36 & 4916 & 2.42 & 1.33   & -0.05 & 0.31        & 0.075          & 0.01  & 0.37        & 0.1            & 0.06  & 0.13                \\
KIC4044238  & 10.012   & 0.2   & 4519 & 2.38 & 1.32   & 0.15  & -0.05       & 0.069          & 0.19  & -0.01       & 0.1            & 0.04  & 0.12                \\
KIC4243796  & 9.712    & 0.11  & 4620 & 2.34 & 1.34   & 0.01  & -0.1        & 0.072          & 0.17  & 0.06        & 0.14           & 0.16  & 0.16                \\
KIC4445711  & 9.484    & -0.32 & 4876 & 2.5  & 1.38   & -0.16 & 0.16        & 0.067          & -0.05 & 0.27        & 0.12           & 0.11  & 0.14                              
\hline
\end{longtable}
}
\twocolumn

\end{appendix}

\end{document}